\documentclass[%
 aip,
 amsmath,amssymb,
preprint,%
]{revtex4-1}
\usepackage{graphicx}
\usepackage{dcolumn}
\usepackage{bm}

\usepackage[utf8]{inputenc}
\usepackage[T1]{fontenc}
\usepackage{mathptmx}
\usepackage{etoolbox}
\usepackage{physics}
\usepackage{amsmath}
\usepackage[caption=false]{subfig}
\usepackage{siunitx}
\usepackage{longtable}

\makeatletter
\def\@email#1#2{%
 \endgroup
 \patchcmd{\titleblock@produce}
  {\frontmatter@RRAPformat}
  {\frontmatter@RRAPformat{\produce@RRAP{*#1\href{mailto:#2}{#2}}}\frontmatter@RRAPformat}
  {}{}
}%
\makeatother
\begin{document}

\preprint{}

\title[]{On The Inclusion of One Double Within CIS and TDDFT}
\author{Vishikh Athavale}
\email{vishikh@sas.upenn.edu}
\author{Hung-Hsuan Teh}
\author{Joseph Subotnik}
\email{subotnik@sas.upenn.edu}
\affiliation{Department of Chemistry, University of Pennsylvania, Philadelphia, Pennsylvania 19104-6323, USA}


\date{\today}

\begin{abstract}
We present an improved approach for generating a set of optimized frontier orbitals (HOMO and LUMO) that minimizes the energy of one double configuration.  We further benchmark the effect of including such a double within a rigorous CIS or parametrized, semi-empirical TDDFT configuration interaction Hamiltonian for a set of test cases.  Although we cannot quite achieve quantitative accuracy, the algorithm is quite robust and routinely delivers an enormous qualitative improvement to standard single-reference electronic structure calculations.
\end{abstract}

\maketitle


\section{\label{sec:introduction}Introduction
}

The need for accurate and computationally inexpensive potential energy surfaces has led to the development of a host of electronic structure methods over the last century. In general, accounting for both static and dynamic correlation requires a careful analysis and a delicate methodology as these effects which add together (in one sense) cannot be disentangled\cite{dynStatCorrelation1,dynStatCorrelation2,dynStatCorrelation3} (in another sense). On the one hand, for chemical processes that involve bond making/breaking, electronic or energy transfer, static correlation is crucial, as one will strongly mix the ground state with a few excited configurations; the true eigenstates of the Hamiltonian will be combinations of a few nearly degenerate determinants. On the other hand, dynamic correlation describes the weak mixing of one electronic state with many other electronic states; one usually imagines the MP2 energy \cite{mp2} as a conventional method to recover dynamical correlation correction.  

Nowadays, for the most part, if one wishes to account for dynamic correlation, the most common (inexpensive) approach is to use Density Functional Theory  (DFT) and Time-Dependent Density Functional Theory (TDDFT). The successes of DFT are almost uncountable\cite{dft1,dft2} even if one must always be hesitant about choosing a functional\cite{dft_functional} and overfitting is clearly a problem\cite{dft_overfitting}.  And yet, despite the formal foundation of DFT as an exact theory (in principle), in practice there is no question that DFT and TDDFT do not recover static correlation correctly. For instance, consider the case of conical intersections\cite{conint1,conint2,conint3,matsikaConical1}, where two or more electronic states become degenerate. Simulating conical intersections is of great interest given that they are known to mediate many photochemical processes such as internal conversion\cite{conint_internalConversion1,conint_internalConversion2}, charge transfer\cite{conint_ct1,conint_ct2}, and isomerization\cite{conint_iso}. For conical intersections between the ground state ($S_0$) and the first excited state ($S_1$), conventional electronic structure methods (including Hartree-Fock (HF) combined with Configuration Interaction Singles (CIS) and DFT/TDDFT) recover the wrong dimensionality of the seam.
More specifically, if a system has $N$ nuclear degrees of freedom, the conical intersection seam should have dimension $N-2$, but  HF/CIS and adiabatic DFT/TDDFT theory both predict a seam of dimension $N-1$ on account of Brillouin's theorem\cite{tddft_failure}. In the end, it is clear that even if we are prepared to use a DFT Hamiltonian as a semi-empirical approximation to accommodate dynamic correlation, such an ansatz must still be corrected to accommodate static correlation. In other words, if we want to introduce a semi-empirical Hamiltonian to treat problems with bond-making/breaking, we require a DFT/TDDFT ansatz with a more balanced treatment of ground state and excited states.

Now, even if we are ready to use the DFT Fock operator and the TDDFT linear response operator as matrix elements of a large configuration interaction Hamiltonian, there are many possibilities for including static correlation. For instance, within a rigorous wave function picture, static correlation is most easily treated with multi-reference methods whereby one optimizes the energy of one many-body state (or a collection of many-body states) by both diagonalizing a large matrix and optimizing orbitals at the same time. 
To that end,  \citeauthor{Gagliardi2017}\cite{Gagliardi2017,gagliardi_lif} have proposed multiconfigurational pair-DFT where a reference wave function of multiconfigurational nature is used for obtaining the classical components of the total energy (kinetic energy, electron-nuclear attraction, and classical electron-electron interactions), which is then combined with the energy from a density functional for the calculation of the non-classical parts of the total energy (such as the exchange-correlation energy). 
Semiempirical DFT/MRCI methods were proposed by \citeauthor{dft_mrci_grimme}\cite{dft_mrci_grimme} and allow calculation of excited states of large molecules\cite{dft_mrci_2}. Recently, quite a few methods have been proposed to rectify the specific failure of DFT/TDDFT\cite{ppRPA,reks,roks,ocdft} so as to recover static correlation and the correct topology of a conical intersection, while retaining the fundamental density functional formulation. For example, Truhlar {\em et al} proposed introducing a non-zero coupling between the DFT ground state and single-excitation  TDDFT response states by artificially using two different functionals within a DFT/TDA calculation ( the so-called  DF-TDA\cite{df-tda} and  CIC-TDA\cite{cic-tda} methods). Spin flip DFT methods\cite{sfdft1,sfdft2} that start from a high-spin triplet state as a reference and produce the ground and excited states using a spin-flip excitation operator have shown strong success in describing conical intersections\cite{sf_conint1,spinFlip_conint}. Through the route presented by constrained DFT\cite{cdft}, one can calculate diabatic representations and excitation energies in the vicinity of conical intersections\cite{cdft-ci}. 
Recently,  \citeauthor{qedft}\cite{qedft} have put forth the QE-DFT method that calculates excited state energies starting from a system deficient by one electron.
pp-RPA methods, which start from a system with a reference determinant of $N\pm2$ electrons to get excitations for an $N$ electron system, have been shown to be useful in accounting for static and dynamic correlation effects. A variant of this method, hh-TDA, has been further refined\cite{hhTDA} with scaled DFT matrix elements in the response kernel to deliver better vertical excitation energies. This list is not exhaustive.\cite{vMSDFT,casDFT}
In short, there are many approaches today that seek to achieve a more balanced treatment of static and dynamic correlation effects within a DFT/TDDFT framework, usually with a semi-empirical flavor.

From our perspective, unfortunately, many of the methods above are not ideal for running nonadiabatic dynamics.   In particular, because some of these methods require a pre-determined set of orbitals in the active space,  dynamics using such methods are prone to user error if the user chooses the wrong set of orbitals (and this error is not corrected until after a long simulation). Moreover, whenever one chooses a set of active orbitals, there is no guarantee that one can generate a smooth set of orbitals (and therefore energies) as a function of nuclear geometry;  it is well known that Complete Active Space Self-Consistent Field (CASSCF) potential energy surfaces can have strange discontinuities\cite{problemCASSCF1,problemCASSCF2,problemCASSCF3}.
In the end,  the ideal merger of DFT theory with static correlation theory would occur in a black-box fashion, such that the user need not choose orbitals, the relevant potential energy surfaces are smooth (or as smooth as possible), and the relevant active space should include the standard DFT ground state.

With this goal in mind, recently we introduced the CIS-1D and TDDFT-1D methods\cite{Teh2019}.  Let us first address CIS-1D.  The CIS-1D method is a straightforward configuration interaction method whose Hamiltonian diagonalizes the following basis functions: the HF determinant $\ket{\psi_0}$, the set of singly excited configurations $\ket{\psi_i^a}$ as in a CIS Hamiltonian, and finally one optimized doubly excited configuration $\ket{\psi_{h\Bar{h}}^{l\Bar{l}}}$, where a pair of electrons are excited from an orbital $\ket{h}$ to $\ket{l}$. As a matter of notation, throughout this manuscript $\{i,j,k,\ldots\}$ denote occupied orbitals, $\{a,b,c,\ldots\}$ denote  virtual orbitals, and $\{p,q,r,\ldots\}$ denote general orbitals. The configuration interaction Hamiltonian in CIS-1D is
\begin{equation}
\label{eq:configHamiltoninan}
    \vb{H}=\begin{pmatrix}
   \epsilon_{HF} & 0 & \mel*{\psi_0}{H}{\psi_{h\Bar{h}}^{l\Bar{l}}}\\
    0 & \mel*{\psi_i^a}{H}{\psi_j^b} & \mel*{\psi_i^a}{H}{\psi_{h\Bar{h}}^{l\Bar{l}}} \\
    \mel*{\psi_0}{H}{\psi_{h\Bar{h}}^{l\Bar{l}}} & \mel*{\psi_{h\Bar{h}}^{l\Bar{l}}}{H}{\psi_i^a} & \mel*{\psi_{h\Bar{h}}^{l\Bar{l}}}{H}{\psi_{h\Bar{h}}^{l\Bar{l}}}
    \end{pmatrix}.
\end{equation}

Second, let us address TDDFT-1D.  To motivate the approach below, note that CIS is analogous to TDDFT within the Tamm-Dancoff Approximation (TDA)\cite{tda1,tda2}-- in so far as both can be implemented by diagonalizing a very similar configuration interaction Hamiltonian of singly excited  determinants.  With this analogy in mind, we fashioned TDDFT-1D as a semi-empirical, parameterized approach for generating excited states in close analogy to CIS-1D.  In particular, the {\em ad hoc} formalism underlying TDDFT-1D was inspired by Maitra et al\cite{Maitra2004} who advocated calculating states of double excitation character within linear response: just as for CIS-1D, to account for static correlation,  we include one doubly excited basis function within  an effective configuration interaction Hamiltonian built on top of a set of Kohn-Sham (KS) reference orbitals. One might view this approach as a continuation of other semi-empirical DFT formulations, such as DFT/MRCI\cite{dft_mrci_grimme}. The goal of TDDFT-1D was to generate (i) vertical excitation energies close to the TDDFT energies but also (ii) smooth $S_1/S_0$ crossings.  In Ref.~\citenum{Teh2019},  using stilbene as an example, we showed that smooth potential energies can be obtained along the torsional coordinate of the double bond, especially around the avoided crossing (where TDDFT fails), while only minimally changing the equilibrium vertical excitation energy. Furthermore, for the conical intersection between the $S_0$ and $S_1$ state of the water molecule, TDDFT-1D recovered the correct topology as well. 

Despite the aforementioned successes, over the past few months, we have run TDDFT-1D and CIS-1D calculations for a host of molecules and found that the algorithm presented in Ref.~\citenum{Teh2019} for finding the optimal double excited configuration does not always converge, and the results were not necessarily stable.  We have also noticed that there were occasions where our semi-empirical TDDFT-1D Hamiltonian yielded smooth $S_1/S_0$ crossings at the cost of less accurate vertical excitation energies.  To that end, the objectives of this manuscript are threefold. First, we will present a new, robust algorithm for calculating the minimum energy doubly excited configuration that is nearly guaranteed to converge (and thus isolate the correct optimized frontier orbitals: the HOMO and the LUMO). Second, we will fine-tune the parameters of our semi-empirical configuration interaction Hamiltonian so as to give a well-balanced treatment of both (i) excitation energies (ii) smooth $S_1/S_0$ crossings. 
Finally, to convince the reader that such a balance can be achieved, we will benchmark the CIS-1D/TDDFT-1D approach on a set of representative examples where one will quickly be able to assess the relative strengths (or weaknesses) of the ansatz.  Overall, our finding is that though the method is not quite quantitatively accurate, the inclusion of a single double clearly offers an outstanding, qualitative correction to single-reference methods.

\section{Theory}\label{theory}
The TDDFT-1D method hinges on selecting one doubly excited configuration to account for static electron correlation. We choose this doubly excited configuration to be the excitation of a pair of electrons from the HOMO $\ket{h}$ to the LUMO $\ket{l}$. In doing so we seek the optimal $\ket{h}$ and $\ket{l}$ orbitals, chosen from the occupied and the virtual space respectively, that minimize the energy for the double excitation. This doubly excited configuration is denoted as $\ket{\psi_{h\bar{h}}^{l\bar{l}}}$. In Ref.~ \citenum{Teh2019}, the optimized $\ket{h}$ and $\ket{l}$ orbitals were isolated by self-consistently finding occupied-occupied and virtual-virtual unitary matrices that independently rotated the occupied and virtual subspaces so as to minimize the energy of the double excitation $E_d(=\mel*{\psi_{h\Bar{h}}^{l\Bar{l}}}{H}{\psi_{h\Bar{h}}^{l\Bar{l}}})$. While we won't repeat the procedure here, it is important to emphasize that the minimization procedure effectively used information from the first derivative of the $E_d$ (as differentiated with respect to orbital rotations). Unfortunately, after applying the procedure in Ref.~\citenum{Teh2019} for bigger molecules,  we have found that the method has shortcomings (see Section \ref{discussion} below). To that end, we will now present an improved algorithm for computing optimized orbitals which is based on a Newton-Raphson optimization technique.

The energy of the doubly excited configuration $E_d$ is a function of the ``inactive'' (non-frontier) orbitals as well as the $\ket{h}$ and $\ket{l}$ orbitals. Starting from an HF reference state, it is straightforward to show that 
\begin{align}
 E_d&= \mel*{\psi_{h\Bar{h}}^{l\Bar{l}}}{H}{\psi_{h\Bar{h}}^{l\Bar{l}}} =E_0 - 2f_{hh}+2f_{ll}+(hh|{h}{h})+(ll|{l}{l})+2(hl|lh)-4(hh|ll).
\end{align}
 Here, $f_{pq}$ are the elements of the Fock matrix, and terms of the form $(pq|rs)$ stand for two-electron integrals in the chemist's notation. $E_0$ is the ground state energy.
Within CIS-1D/TDDFT-1D, the optimized orbitals $\ket{h}$ and $\ket{l}$ are generated from independent rotations of the orbitals in the occupied and virtual spaces, respectively. We emphasize that such rotations keep the energy of the ground state unchanged, and so one can effectively consider $E_d$ to be a function of $\ket{h}$ and $\ket{l}$ only. Thus, if the relevant rotation matrices have the form $e^{\bm{\Theta}}$ (where $\bm{\Theta}$ is an anti-symmetric matrix), we must compute only the optimized HOMO and the LUMO so that the $\Theta$ matrix takes the following simple form: 
\begin{equation}
\label{theta_mat}    
\Theta =\left(\begin{array}{@{}c|c@{}}
     \begin{matrix} 
    0 & \dots & 0 & \theta_{1h} \\
    0 & \dots & 0 & \theta_{2h} \\
    \vdots & \ddots & & \vdots\\
    0 & \dots & 0 & \theta_{h-1,h} \\
    -\theta_{1h} & \ldots & -\theta_{h-1,1} & 0 
    \end{matrix}
    & \mbox{\normalfont\Large\bfseries 0} \\
    \hline
    \mbox{\normalfont\Large\bfseries 0} &
    \begin{matrix}
    0 & -\theta_{l+1,l}&\ldots&-\theta_{ln}\\
    \theta_{l+1,l} & 0 &\ldots&0\\
    \vdots&\vdots&\ddots\\
    \theta_{nl} & 0 & \ldots &0
    \end{matrix}
    \end{array}\right).
\end{equation}
For infinitesimally small rotations, the orbitals $\ket{h}$ and $\ket{l}$ after rotation are
\begin{subequations}
\begin{align}
    \ket{\Tilde{h}}&=\ket{h}-\sum_i\theta_{ih}\ket{i}-\sum_i\frac{\theta_{ih}^2}{2}\ket{h},\\
\ket{\Tilde{l}}&=\ket{l}-\sum_a\theta_{al}\ket{i}-\sum_a\frac{\theta_{al}^2}{2}\ket{l}.
\end{align}
\end{subequations}

Using these so-called ``exponential'' coordinates, $E_d$ (which is now a function of $\ket{\Tilde{h}}$ and $\ket{\Tilde{l}}$) can be written as a function of $\{\theta_{ih}\}$ and $\{\theta_{al}\}$. We will now apply the Newton-Raphson optimization as a function of the variables $\{\{\theta_{ih}\}$, $\{\theta_{al}\}\}$.  According to such an approach, we will take steps in $\Theta$ that are informed from the gradient and Hessian of $E_d$. These quantities are given by
\begin{subequations}
   \label{ed_differentiate}
   \begin{align}
    \pdv{E_d}{\theta_{ih}}&=4\left(f_{ih}-(ih|hh)+2(ih|ll)-(il|lh)\right)\\
    \pdv[2]{E_d}{\theta_{ih}}{\theta_{jh}}&= 4\delta_{ij}\left(f_{hh}-(hh|hh)+2(hh|ll)-(hl|lh)\right)\nonumber
       \\ &\quad -4\left(f_{ij}-2(hi|hj)+(hh|ij)+2(ij|ll)-(il|lj)\right)\\
       \pdv{E_d}{\theta_{al}}&=4\left(f_{al}-(al|ll)+2(al|hh)-(ah|hl)\right)\\
    \pdv[2]{E_d}{\theta_{al}}{\theta_{bl}}&= -4\delta_{ab}\left(f_{ll}-(ll|ll)+2(hh|ll)-(hl|lh)\right)\nonumber
       \\ &\quad +4\left(f_{ab}-2(hh|ab)+(hb|ah)+2(la|bl)-(ll|ab)\right)\\
       \pdv[2]{E_d}{\theta_{al}}{\theta_{ih}}&=4(ia|lh)+4(il|ah)-16(ih|al).
\end{align}
\end{subequations}

We may now summarize our algorithm for computing the optimized frontier molecular orbitals:
\begin{enumerate}
    \item Solve the self-consistent field equations to obtain the cannonical HF molecular orbital (MO) coefficient matrix $\vb{C}.$
    \item Calculate the gradient vector 
    $\vb{g}=\left( \begin{array}{c} \pdv{E_d}{\theta_{ih}} \\\pdv{E_d}{\theta_{al}}\end{array}\right)$
    and the Hessian $\mathcal{H}.$
    \item Solve for $\theta_{ih}$ and $\theta_{al}$ using the equation $\mathcal{H} \bm{\theta}=-\vb{g}$. Here $\bm{\theta}=\left(\begin{array}{c}
      \theta_{ih} \\ \theta_{al}
    \end{array}\right).$
    \item Construct the $\bm{\Theta}$ matrix as described in eq \ref{theta_mat}.
    \item Obtain the new MO coefficients $\Tilde{\vb{C}}=\vb{C} e^{\bm{\Theta}}$.
    \item If $|\vb{\Tilde{C}}-\vb{C}|$ is below a threshold then the optimization has converged. Else go to setp 2 with $\vb{C}=\vb{\Tilde{C}}.$
   \end{enumerate} 
    Note that, for the algorithm above, the Hessian is only of size $N - 2$, where $N$ is the size of the atomic basis. For the most part, inversion of $\mathcal{H}$ carries a minimal cost.
    
    Extending the method above to work with TDDFT-1D is also straightforward. One works with the KS orbitals as if they were real orbitals and one defines a key double excitation of the KS ground state wave function from the HOMO to the LUMO. The energy of this doubly excited configuration is 
    \begin{equation}
      E_d=E[\rho_d]=2\sum_{i}{}^{'} h_{ii}+2\sum_{ij}{}^{'} (ii|jj) - c_{HF}(ij|ji) + (1-c_{HF})E_{xc}[\rho_d]  
    \end{equation}
     The electron density obtained by exciting a pair of electrons from orbital $\ket{h}$ to $\ket{l}$ is denoted by $\rho_d$. $h_{ii}$ refers to the one-electron interaction terms. The summation in the expression is primed to indicate that the summation is over the orbitals $1,2,\ldots,h-1,l$ (orbital $\ket{h}$ is omitted and $\ket{l}$ is included). For a hybrid xc functional $c_{HF}$ is the factor of the HF exchange. $E_{xc}[\rho]$ is the xc energy functional for a given density $\rho$. Just as for the CIS-1D case,  in order to find the optimized doubly excited configuration we rotate the KS orbitals using a rotation operator of the form $e^{\bm{\Theta}}.$ As compared with the expressions for the gradient and Hessian above (Eq. \ref{ed_differentiate}), besides the factor of $c_{HF}$, the gradient and Hessian have the following additional terms from differentiating the xc functional:
     \begin{align}
    \pdv{E_{xc}[\rho_d]}{\theta_{ih}}&=\int \phi_i(\vb{r})\eval{\pdv{E_{xc}}{\rho(\vb{r})}}_{\rho=\rho_d}\phi_h(\vb{r})\dd\vb{r}\\
    \pdv[2]{E_{xc}[\rho_d]}{\theta_{ih}}{\theta_{jh}}&= \int \phi_h(\vb{r})\phi_i(\vb{r})\eval{\pdv[2]{E_{xc}}{\rho(\vb{r})}{\rho(\vb{r'})}}_{\rho=\rho_d}\phi_h(\vb{r'})\phi_j(\vb{r'})\dd\vb{r}\dd\vb{r'}.
    \label{eq:dft_ed_diff}
\end{align}

\subsection{Choice of parameters for TDDFT-1D}
 For the case of  CIS-1D, there is a rigorous Hamiltonian $H$ and calculating the matrix elements  $\mel*{\psi_0}{H}{\psi_{h\Bar{h}}^{l\Bar{l}}}$ and $\mel*{\psi_i^a}{H}{\psi_{h\Bar{h}}^{l\Bar{l}}}$ is straightforward: One simply sets:
  \begin{subequations}
  \label{eq:cis1d-offdiag}
  \begin{align}
\mel*{\psi_0}{H}{\psi_{h\Bar{h}}^{l\Bar{l}}} &= (hl|hl) \label{eq:beta}\\
\mel*{\psi_i^a}{H}{\psi_{h\Bar{h}}^{l\Bar{l}}} &= \delta_{ih}(al|hl) - \delta_{al}(hl|hi) \label{eq:alpha}   
  \end{align}
  \end{subequations}

 However, for a semi-empirical TDDFT-1D approach, the KS-DFT  reference is not a true wave function, TDA is itself an approximation to a true TDDFT response theory, and there is no rigorous or unique way to estimate the coupling between the ground state and the double or the couplings between the double and the singles. Note that this state of affairs has led to  a large discussion of how to construct TD-DFT derivative couplings in the past.\cite{tddft_dc1,tddft_dc2,tddft_dc3,tddft_dc4,tddft_dc5,tddft_dc6} Most recently, in   Ref.\citenum{Teh2019}, we simply approached the TDDFT-1D couplings following the CIS-1D approach, i.e. we presumed that the KS reference configurations could be treated as legitimate configuration interaction wave functions and used the exact Hamiltonian as the operator to be diagonalized. Below, however,  we will show that (empirically) a far better approach is to parameterize such coupling matrix elements by multiplying with a scaling factor; otherwise, the raw coupling matrix elements in Eq.~\ref{eq:cis1d-offdiag} (which involve exchange) are simply too large in practice.


Let us now be more precise. Below, we will multiply the singles-double couplings (in Eq.~\ref{eq:alpha}) by a factor $\alpha$; we will multiply the ground-double coupling (in Eq.~\ref{eq:beta}) by a factor $\beta$. On the one hand, a key requirement of TDDFT-1D is that, as far as equilibrium geometry vertical excitation energies are concerned, one does not want to strongly perturb the TDA energies.  As such, one would like $\alpha$ and $\beta$ to be as small as possible. On the other hand, another requirement for TDDFT-1D is smooth $S_1/S_0$ crossings. The factor $\beta$ that couples the KS determinant and the doubly excited configuration must be large enough to lend the doubly excited character to the reference KS wave-function.  After testing on a reasonably large set of data over a range from 0 to 1, the values of $\alpha=0.5$ and $\beta=0.75$ were found to be optimal for the B3LYP functional. Results will be presented below for a benchmarking set of data. 

\section{Results}
We have implemented the algorithm within a development version of the Q-Chem 5.3 software\cite{qchem} above so as to apply the CIS-1D and the TDDFT-1D formalisms to reasonably sized molecules. We will now discuss our findings as far as balancing (i) accurate equilibrium vertical excitation energies vs. (ii) smooth $S_1/S_0$ crossings.

\subsection{Equilibrium Vertical Excitation Energies}\label{benchmark_section}
 To begin our investigation, we benchmark TDDFT-1D vertical excitation energies on a set of 28 medium-sized organic molecules originally  investigated by Thiel and coworkers\cite{thiel_benchmark}. We wish to verify that TDDFT-1D does not significantly change the vertical excitation energies of TDDFT/TDA; we will further compare and benchmark the reference excitation energies against those calculated from reasonably high-level wave function methods such as CASPT2 and CC3 in Ref.~\citenum{thiel_benchmark}. We use the B3LYP functional since this gave the best performance in the original benchmark study and use the TZVP basis set as in Ref.~\citenum{thiel_benchmark}. Excitations of the type $\pi\rightarrow\pi^*$, $n\rightarrow\pi^*$, and $\sigma\rightarrow\pi^*$ were originally identified in the test molecules for a TDDFT/TDA calculation. In this work, we identified the same states in a TDDFT-1D calculations by carefully looking at the states and comparing their transition dipole moments and oscillator strengths. 

In Table~\ref{benchmark_s0s1_table} we list the vertical excitation energies obtained from a CASPT2, CC3, TDDFT/TDA, TDDFT-1D  ($\alpha=0.5$, $\beta=0.75$), and TDDFT-1D ($\alpha=1$, $\beta=1$) calculation. The absolute mean deviation (AMD) between TDDFT-1D (with or without empirical scaling factors) and TDDFT/TDA is small ($\leq$\SI{0.18}{\eV}),  much smaller than the AMD between CASPT2 and TDDFT/TDA (\SI{0.29}{\eV}) or between CC3 and TDDFT/TDA (\SI{0.29}{\eV}). Thus, it appears that TDDFT-1D does not move vertical energies appreciably from TDDFT/TDA. With the emperical scaling ($\alpha=0.5$, $\beta=0.75$) TDDFT-1D performs at the same level in the benchmarking test as TDDFT/TDA (both have an AMD of \SI{0.29}{\eV} from the high level calculations) and TDDFT-1D without scaling only does a little worse (with an AMD of \SI{0.32}{\eV}). 
 

Lastly, it is instructive to look at the overlap between the $S_1$ state obtained from TDDFT-1D and that from TDDFT/TDA: do these states match up in terms of character and overall wave function shape? Among the molecules considered in Table \ref{benchmark_s0s1_table}, we find that butadiene, octatetraene, and cyclopropene have states where the ordering between the first and second excited states is switched, as there is a considerable double excitation character in the $S_1$ state. Interestingly, in the DFT/MRCI result in the original benchmarking study in Ref.~\citenum{thiel_benchmark} there is a significant contribution from a HOMO to LUMO double excitation to the $^1A_g$ state in these three molecules. This led to the switching of the energy ordering of the $^1B_u$ state and the $^1A_g$ state (when compared with the TDDFT ordering) for octatetraene and cyclopropene, making the latter lower in energy. Multi-reference calculations have verified the predominance of HOMO to LUMO double excitation contribution to excited states in polyenes\cite{double_excite}.

\begin{longtable}{cccccccc}
\caption[]{Vertical excitation energies (eV) of 28 small-to-medium organic molecules as calculated by  TDDFT/TDA/B3LYP, TDDFT-1D/B3LYP with  scaling parameters {($\alpha=1.0$, {$\beta=1.0$)}}, and scaling parameters {($\alpha=0.5$, {$\beta=0.75$)}}  in the basis set TZVP compared to the CASPT2 and CC3 energies from literature\cite{thiel_benchmark}. Note that, as far as excitation energies are concerned,  the absolute mean deviation (AMD) and root mean square deviation (RMSD) between TDDFT-1D and TDDFT/TDA is smaller than the AMD between the TDDFT/TDA results and the benchmark wave function (CASPT2/CC3) results.}
\label{benchmark_s0s1_table}\\
%
\hline
\textbf{Molecule} &
  \textbf{\begin{tabular}[t]{@{}c@{}}TDDFT/TDA \\ State no.\end{tabular}} &
  \textbf{\begin{tabular}[t]{@{}c@{}}TDDFT-1D \\ State no.\end{tabular}} &
  \textbf{TDDFT/TDA} &
  \textbf{\begin{tabular}[t]{@{}c@{}}TDDFT-1D\\ $\mathbf{(1.0,1.0)}$\end{tabular}} &
  \textbf{\begin{tabular}[t]{@{}c@{}}TDDFT-1D\\ $\mathbf{(0.5,0.75)}$\end{tabular}} &
  \textbf{CASPT2} &
  \textbf{CC3} \\ \hline
Ethene             & 3  & 3  & 8.30 & 9.08 & 8.75 & 8.54 & 8.37  \\
E-butadiene        & 2  & 2  & 6.83 & 6.89 & 6.59 & 6.62 & 6.77  \\
                   & 1  & 1  & 6.26 & 5.69 & 6.52 & 6.47 & 6.58  \\
All-E-hexatriene   & 2  & 2  & 5.71 & 5.71 & 5.45 & 5.42 & 5.72  \\
                   & 1  & 1  & 5.17 & 4.46 & 5.26 & 5.31 & 5.58  \\
All-E-octatetraene & 2  & 2  & 4.85 & 4.94 & 4.70 & 4.64 & 4.97  \\
                   & 1  & 1  & 4.46 & 3.64 & 4.38 & 4.70 & 4.94  \\
Cyclopropene       & 2  & 2  & 6.63 & 7.21 & 6.96 & 7.06 & 7.10  \\
                   & 1  & 1  & 6.48 & 7.06 & 6.81 & 6.76 & 6.90  \\
Cyclopentadiene    & 3  & 2  & 6.54 & 5.83 & 6.39 & 6.31 & 6.61  \\
                   & 12 & 12 & 8.66 & 9.07 & 8.89 & 8.52 & 8.69  \\
                   & 1  & 1  & 5.40 & 5.79 & 5.63 & 5.51 & 5.73  \\
Norbornadiene      & 1  & 1  & 4.95 & 5.10 & 5.03 & 5.34 & 5.64  \\
                   & 5  & 5  & 7.24 & 7.39 & 7.32 & 7.45 & 7.71  \\
                   & 2  & 2  & 5.55 & 5.69 & 5.63 & 6.11 & 6.49  \\
                   & 4  & 4  & 7.01 & 7.16 & 7.09 & 7.32 & 7.64  \\
Benzene            & 2  & 2  & 6.33 & 6.67 & 6.52 & 6.42 & 6.68  \\
                   & 1  & 1  & 5.43 & 5.76 & 5.62 & 5.04 & 5.07  \\
                   & 5  & 5  & 7.65 & 7.99 & 7.84 & 7.13 & 7.45  \\
Naphthalene        & 4  & 4  & 6.20 & 5.83 & 6.30 & 5.87 & 5.98  \\
                   & 2  & 2  & 4.55 & 4.78 & 4.68 & 4.77 & 5.03  \\
                   & 1  & 1  & 4.46 & 4.69 & 4.58 & 4.24 & 4.27  \\
                   & 3  & 3  & 5.59 & 5.82 & 5.71 & 5.99 & 6.07  \\
                   & 8  & 8  & 6.63 & 6.77 & 6.67 & 6.47 & 6.79  \\
Furan              & 2  & 1  & 6.76 & 6.38 & 6.76 & 6.50  & 6.62  \\
                   & 10 & 10 & 8.83 & 9.15 & 9.02 & 8.17 & 8.53  \\
                   & 1  & 2  & 6.55 & 6.95 & 6.77 & 6.39 & 6.60   \\
Pyrrole            & 3  & 2  & 6.60 & 6.44 & 6.67 & 6.31 & 6.40   \\
                   & 12 & 12 & 8.49 & 8.85 & 8.67 & 8.17 & 8.17  \\
                   & 4  & 4  & 6.75 & 7.13 & 6.95 & 6.33 & 6.71  \\
Imidazole          & 3  & 3  & 6.70 & 7.02 & 6.90 & 6.19 & 6.58  \\
                   & 4  & 4  & 7.21 & 7.33 & 7.35 & 6.93 & 7.10   \\
                   & 11 & 11 & 8.64 & 8.94 & 8.83 & 8.16 & 8.45  \\
                   & 2  & 2  & 6.49 & 6.85 & 6.69 & 6.81 & 6.82  \\
                   & 7  & 7  & 7.51 & 7.87 & 7.71 & 7.90  & 7.93  \\
Pyridine           & 4  & 4  & 6.56 & 6.57 & 6.58 & 6.39 & 6.85  \\
                   & 9  & 9  & 8.00 & 8.06 & 8.04 & 7.46 & 7.70   \\
                   & 3  & 3  & 5.56 & 5.63 & 5.60 & 5.02 & 5.15  \\
                   & 8  & 8  & 7.78 & 7.85 & 7.82 & 7.27 & 7.59  \\
                   & 1  & 1  & 4.85 & 4.92 & 4.89 & 5.17 & 5.05  \\
                   & 2  & 2  & 5.12 & 5.17 & 5.15 & 5.51 & 5.50   \\
Pyrazine           & 6  & 6  & 6.79 & 6.81 & 6.80 & 6.89 & 7.07  \\
                   & 10 & 10 & 8.23 & 8.25 & 8.24 & 7.79 & 8.06  \\
                   & 3  & 3  & 5.52 & 5.54 & 5.53 & 4.85 & 5.02  \\
                   & 9  & 9  & 8.09 & 8.11 & 8.10 & 7.66 & 8.05  \\
                   & 2  & 2  & 4.70 & 4.72 & 4.71 & 4.70  & 5.05  \\
                   & 5  & 5  & 6.39 & 6.41 & 6.40 & 6.41 & 6.75  \\
                   & 4  & 4  & 5.62 & 5.64 & 5.63 & 5.68 & 5.74  \\
                   & 1  & 1  & 4.02 & 4.04 & 4.03 & 4.12 & 4.24  \\
Pyrimidine         & 6  & 6  & 6.85 & 6.89 & 6.87 & 6.63 & 7.06  \\
                   & 8  & 8  & 7.96 & 8.00 & 7.98 & 7.21 & 7.74  \\
                   & 4  & 4  & 5.81 & 5.85 & 5.83 & 5.24 & 5.36  \\
                   & 9  & 9  & 8.27 & 8.26 & 8.29 & 7.64 & 8.01  \\
                   & 1  & 1  & 4.30 & 4.33 & 4.32 & 4.44 & 4.50   \\
                   & 2  & 2  & 4.62 & 4.65 & 4.64 & 4.80  & 4.93  \\
Pyridazine         & 4  & 4  & 5.68 & 5.71 & 5.70 & 5.18 & 5.22  \\
                   & 10 & 10 & 8.07 & 8.09 & 8.09 & 7.62 & 7.82  \\
                   & 6  & 6  & 6.69 & 6.72 & 6.71 & 6.31 & 6.93  \\
                   & 9  & 9  & 7.74 & 7.78 & 7.76 & 7.29 & 7.55  \\
                   & 2  & 2  & 4.21 & 4.23 & 4.22 & 4.31 & 4.49  \\
                   & 3  & 3  & 5.48 & 5.43 & 5.48 & 5.77 & 5.74  \\
                   & 1  & 1  & 3.66 & 3.70 & 3.68 & 3.78 & 3.92  \\
                   & 5  & 5  & 6.11 & 6.14 & 6.13 & 6.52 & 6.41  \\
ss-triazine        & 6  & 6  & 7.29 & 7.29 & 7.29 & 7.25 & 7.41  \\
                   & 5  & 5  & 6.17 & 6.17 & 6.17 & 5.79 & 5.71  \\
                   & 10 & 10 & 8.27 & 8.27 & 8.27 & 7.50  & 8.04  \\
                   & 1  & 1  & 4.46 & 4.46 & 4.46 & 4.60  & 4.78  \\
                   & 2  & 2  & 4.57 & 4.57 & 4.57 & 4.66 & 4.76  \\
                   & 3  & 3  & 4.57 & 4.57 & 4.57 & 4.70  & 4.81  \\
                   & 8  & 8  & 7.56 & 7.56 & 7.56 & 7.71 & 7.80   \\
s-tetrazine        & 2  & 2  & 3.53 & 3.56 & 3.55 & 3.51 & 3.79  \\
                   & 4  & 4  & 5.09 & 5.12 & 5.11 & 5.50  & 5.46  \\
                   & 3  & 3  & 4.82 & 4.85 & 4.83 & 4.73 & 4.97  \\
                   & 5  & 5  & 5.32 & 5.24 & 5.23 & 5.20  & 5.34  \\
                   & 1  & 1  & 2.33 & 2.35 & 2.34 & 2.29 & 2.53  \\
Formaldehyde       & 1  & 1  & 3.89 & 3.91 & 3.90 & 3.99 & 3.95  \\
                   & 3  & 3  & 8.97 & 8.98 & 8.97 & 9.14 & 9.18  \\
                   & 5  & 5  & 9.41 & 9.36 & 9.40 & 9.32 & 10.45 \\
Acetone            & 1  & 1  & 4.36 & 4.37 & 4.36 & 4.44 & 4.40   \\
                   & 9  & 9  & 9.33 & 9.33 & 9.33 & 9.31 & 9.65  \\
                   & 7  & 7  & 8.65 & 8.66 & 8.65 & 9.27 & 9.17  \\
p-benzoquinone     & 2  & 2  & 2.61 & 2.61 & 2.61 & 2.77 & 2.85  \\
                   & 1  & 1  & 2.46 & 2.47 & 2.47 & 2.76 & 2.75  \\
                   & 4  & 4  & 5.19 & 5.20 & 5.20 & 5.28 & 5.62  \\
                   & 3  & 3  & 3.84 & 3.85 & 3.84 & 4.26 & 4.59  \\
                   & 11 & 11 & 6.72 & 6.48 & 6.48 & 6.96 & 7.27  \\
                   & 5  & 5  & 5.45 & 5.46 & 5.45 & 5.64 & 5.82  \\
Formamide          & 1  & 1  & 5.57 & 5.51 & 5.58 & 5.63 & 5.65  \\
                   & 6  & 6  & 8.57 & 8.58 & 8.59 & 7.39 & 8.27  \\
Acetamide          & 1  & 1  & 5.58 & 5.55 & 5.59 & 5.69 & 5.69  \\
                   & 4  & 4  & 7.60 & 7.61 & 7.61 & 7.27 & 7.67  \\
Propanamide        & 1  & 1  & 5.61 & 5.58 & 5.61 & 5.72 & 5.72  \\
                   & 4  & 4  & 7.34 & 7.37 & 7.36 & 7.20  & 7.62  \\
Cytosine           & 2  & 2  & 4.81 & 5.06 & 4.95 & 4.67 &       \\
                   & 4  & 4  & 5.57 & 5.83 & 5.72 & 5.53 &       \\
                   & 11 & 11 & 6.73 & 6.99 & 6.88 & 6.40  &       \\
                   & 10 & 10 & 6.61 & 6.85 & 6.75 & 6.97 &       \\
                   & 1  & 1  & 4.76 & 5.03 & 4.91 & 5.12 &       \\
                   & 3  & 3  & 5.14 & 5.40 & 5.28 & 5.53 &       \\
Thymine            & 2  & 2  & 5.22 & 5.59 & 5.43 & 5.06 &       \\
                   & 4  & 4  & 6.11 & 6.31 & 6.27 & 6.15 &       \\
                   & 7  & 7  & 6.49 & 6.86 & 6.70 & 6.53 &       \\
                   & 1  & 1  & 4.70 & 5.08 & 4.91 & 4.95 &       \\
                   & 3  & 3  & 5.81 & 6.19 & 6.02 & 6.38 &       \\
                   & 6  & 6  & 6.22 & 6.59 & 6.43 & 6.85 &       \\
                   & 8  & 8  & 6.70 & 7.08 & 6.91 & 7.43 &       \\
Uracil             & 2  & 2  & 5.39 & 5.75 & 5.60 & 5.23 &       \\
                   & 4  & 4  & 5.99 & 6.14 & 6.14 & 6.15 &       \\
                   & 8  & 8  & 6.67 & 7.03 & 6.87 & 6.74 &       \\
                   & 12 & 12 & 7.63 & 8.00 & 7.83 & 7.42 &       \\
                   & 1  & 1  & 4.63 & 5.00 & 4.83 & 4.91 &       \\
                   & 3  & 3  & 5.75 & 6.12 & 5.95 & 6.28 &       \\
                   & 5  & 5  & 6.14 & 6.50 & 6.34 & 6.98 &       \\
                   & 7  & 7  & 6.65 & 7.02 & 6.85 & 7.28 &       \\
Adenine            & 3  & 3  & 5.39 & 5.61 & 5.52 & 5.20  &       \\
                   & 2  & 2  & 5.16 & 5.28 & 5.26 & 5.29 &       \\
                   & 9  & 9  & 6.51 & 6.71 & 6.62 & 6.34 &       \\
                   & 12 & 12 & 6.89 & 7.06 & 6.99 & 6.64 &       \\
                   & 1  & 1  & 4.98 & 5.22 & 5.11 & 5.19 &       \\
                   & 4  & 4  & 5.64 & 5.87 & 5.76 & 5.96 &       \\ \hline
 & &
  &
  \textbf{TDDFT/TDA} &
  \textbf{\begin{tabular}[t]{@{}c@{}}TDDFT-1D\\ $\mathbf{(1.0,1.0)}$\end{tabular}} &
  \textbf{\begin{tabular}[t]{@{}c@{}}TDDFT-1D\\ $\mathbf{(0.5,0.75)}$\end{tabular}} &
  \textbf{CASPT2} &
  \textbf{CC3} \\ \hline                   
\textbf{TDDFT-1D $\mathbf{(0.5,0.75)}$} &   &   &   &   &   &   &
   \\ \hline
AMD        &    &    & 0.10 & 0.10 &  & 0.29 & 0.29  \\
RMSD             &    &    & 0.13 & 0.18 &  & 0.36 & 0.34  \\ \hline
\textbf{TDDFT/TDA }       &    &    &      &      &      &      &       \\ \hline
AMD              &    &    &  & 0.18 &  & 0.29 & 0.29  \\
RMSD             &    &    &  & 0.26 &  & 0.36 & 0.34  \\ \hline
\textbf{TDDFT-1D $\mathbf{(1.0,1.0)}$}    &    &    &      &      &      &      &       \\ \hline
AMD              &    &    &  &  &  & 0.32 & 0.33  \\
RMSD             &    &    &  &  &  & 0.42 & 0.41  \\ \hline
\end{longtable}

\subsection{Smooth $S_1/S_0$ crossings}
Having addressed whether or not a 1D ansatz changes vertical excitation energies, we  will now address
if and how a 1D ansatz corrects $S_1/S_0$ crossings.

\subsubsection{PYCM}
We begin with the molecule 2-(4-(propan-2-ylidene)-
cyclohexylidene)malononitril), or PYCM, which is asymmetric about the C-C double bond, in contrast to ethylene. PYCM is a donor-acceptor molecule that possesses two low-lying excited states, one of which is of charge transfer (CT) character and the other one has local excitation character\cite{pycm-verhoeven}. Its photoemission spectrum indicates that the CT state fluoresces whereas the non-CT state undergoes a radiationless decay. The twisting of the double bond between the cyclohexane ring and the cyano groups gives rise to an avoided crossing between the $S_1$ and $S_2$ states, which mediates photochemical processes.  Previous work has analyzed the $S_2/S_1$ crossing at angles $\sim\ang{40}$\cite{pycm_xinle_jctc} but has not directly addressed a potential crossing at \ang{90} when the molecule fully distorts (and the double bond breaks). Presumably, one can expect that twisting PYCM should behave something like twisting ethylene.

\begin{figure}
    \includegraphics[scale=.91]{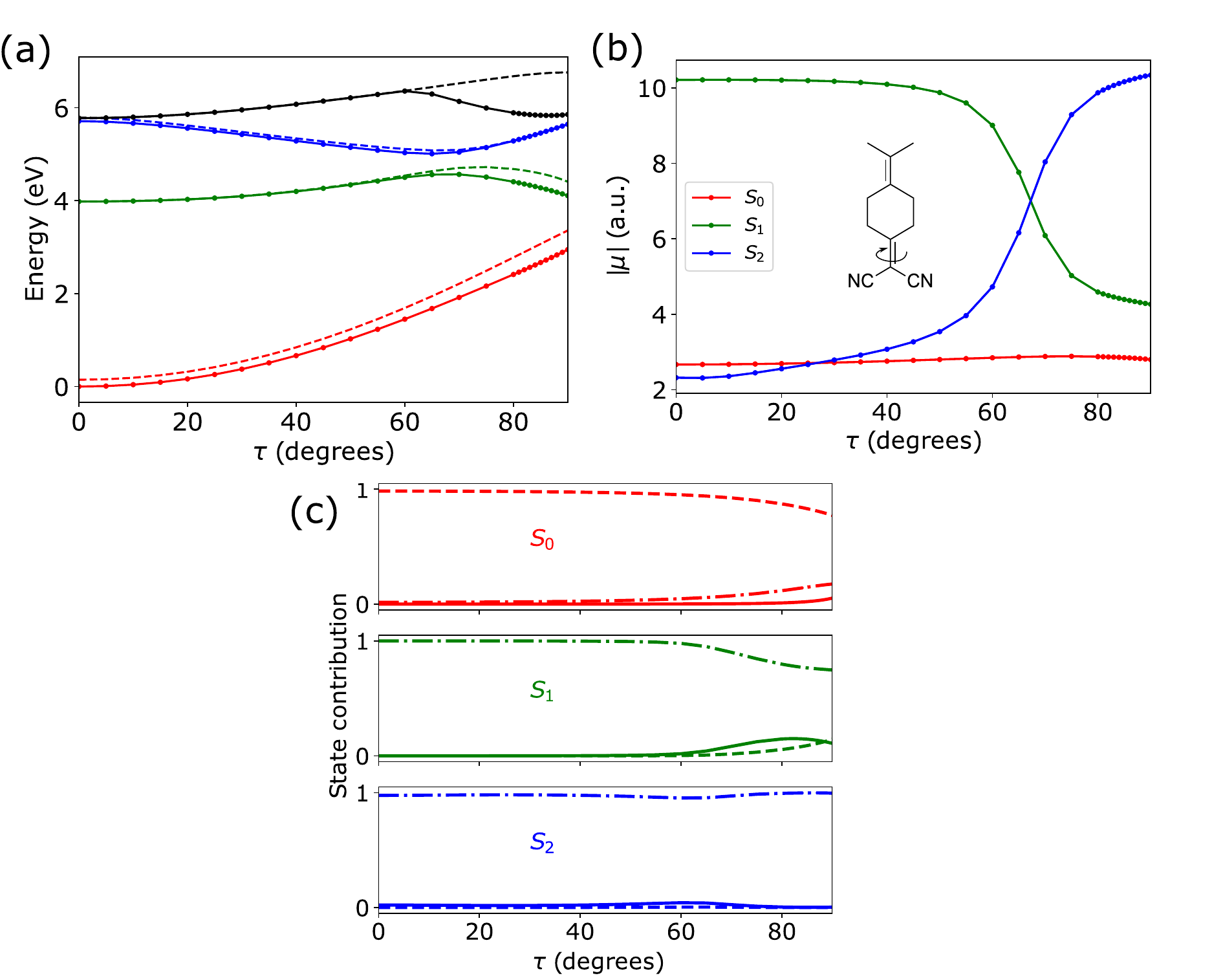}
    \caption{(a)Potential energy curves for the PYCM molecule (left) and (b) dipole moments $|\mu|$ (right) along the torsional angle $\tau$. On the left, the $S_0, S_1,S_2,S_3$ energies for TDDFT-1D B3LYP/6-31G* are plotted using solid lines in red, green, blue, and black respectively. The dashed lines represent the energies for the same states calculated using B3LYP/TDA. The dipole moments of the TDDFT-1D states $S_1$ and $S_2$ cross around \ang{62}. The torsional angle $\tau$ is represented in the inset in the plot on the right. Note that there is no avoided crossing between $S_1/S_0$ states at \ang{90} according to an analysis of their dipole moments. In (c) the contribution of the KS determinant (dashed line), singly excited configurations (dash-dot line), and double (solid line) to each of the TDDFT-1D $S_0,S_1,S_2$ wave functions is plotted. For states $S_0$ and $S_1$ near the breakdown of the C-C double bond (\ang{90}) the contribution of double raises, whereas around the equilibrium geometry (\ang{0}) the contribution is virtually zero. Note that state mixing occurs (and the 1D configuration is important) only around \ang{90}.
    }
    \label{pycm_fig}
\end{figure}

In Fig.~\ref{pycm_fig}, we present the potential energy curves of PYCM starting from its ground state minimum geometry and following along the torsional angle of the double bond. We evaluate the performance of the TDDFT-1D/B3LYP method against that of the TDA/B3LYP using a basis set of 6-31G*. For the TDDFT-1D calculation, the key doubly excited configuration is easily found using the algorithm described above in Sec. \ref{theory}; convergence details will be provided in Sec. \ref{discussion} below.  According to Fig.~\ref{pycm_fig}, TDDFT-1D  yields smooth curves that are compatible with nonadiabatic photodynamics simulations -- meaningful changes to TDDFT/TDA are applied only around \ang{90}. \footnote{Interestingly, if one wishes to apply CCSD/EOM-CCSD to study PYCM photodynamics, one finds the well-known result\cite{hattig_ccsd} that the method is quite discontinuous and fails dramatically around \ang{90}.}   This conclusion is confirmed in Fig.~\ref{pycm_fig}(c) where we plot the contributions of the doubles/singles/KS determinant to the final configuration interaction eigenstates.  

Lastly, in Fig.~\ref{pycm_fig}(b), we also plot the dipole moments of the individual excited states. Note that there is a crossing between $S_2$ and $S_1$--the dipole moments of these states switch character around \ang{67}. However, there is no such crossing between the dipole moments of $S_1$ and $S_0$.  Moreover, although not plotted here, we find that a variety of diabatization schemes (including Boys\cite{boys_subotnik2008}, etc.) do not predict a meaningful diabatic crossing according to this degree of freedom--which has clear consequences for nonadiabatic dynamics.  Presumably,  as for the case of ethylene\cite{ethylene_ci}, there will be a conical intersection between $S_0$ and $S_1$ at some distorted geometry, but no such intersection is present according to the unrelaxed scan in Fig.~\ref{pycm_fig}(a).  In the future, once a TDDFT-1D gradient has been implemented, we will indeed scan for the minimal energy conical intersection. 


\subsubsection{Thiophene}
Our next example is the thiophene molecule whose oligomers have drawn attention as materials useful for energy conversion technology\cite{thiophene_energy1,thiophene_energy2,thiophene_energy3}. The photo-decay for a standard thiophene molecule is dominated by internal conversion through a conical intersection, whereas higher chain oligothiophenes show a significant rate of intersystem crossing (ISC)\cite{thiophene_main}. The photo-deactivation pathway for bithiophene (occurring through ISC as well as through $S_1/S_0$ conical intersection) was investigated in Ref.\citenum{thiophene_main} using EOM-CCSD. Here we investigate the potential energy curves of bithiophene using the TDDFT-1D method.  For this molecule, an optimized ground state geometry at the B3LYP/6-311G** level was reported in Ref.~\citenum{thiophene_main}. Furthermore, there are two $S_1$ minimum structures at the CASSCF/6-31G* level, $S_1$-min-$a$ and $S_1$-min-$b$, which are closed- and open-ring structures respectively. There are some low-lying conical intersection geometries between $S_0$ and $S_1$ identified at the CASSCF level in Ref.~\citenum{thiophene_main}.

\begin{figure}
    \subfloat{
    \includegraphics[scale=0.52]{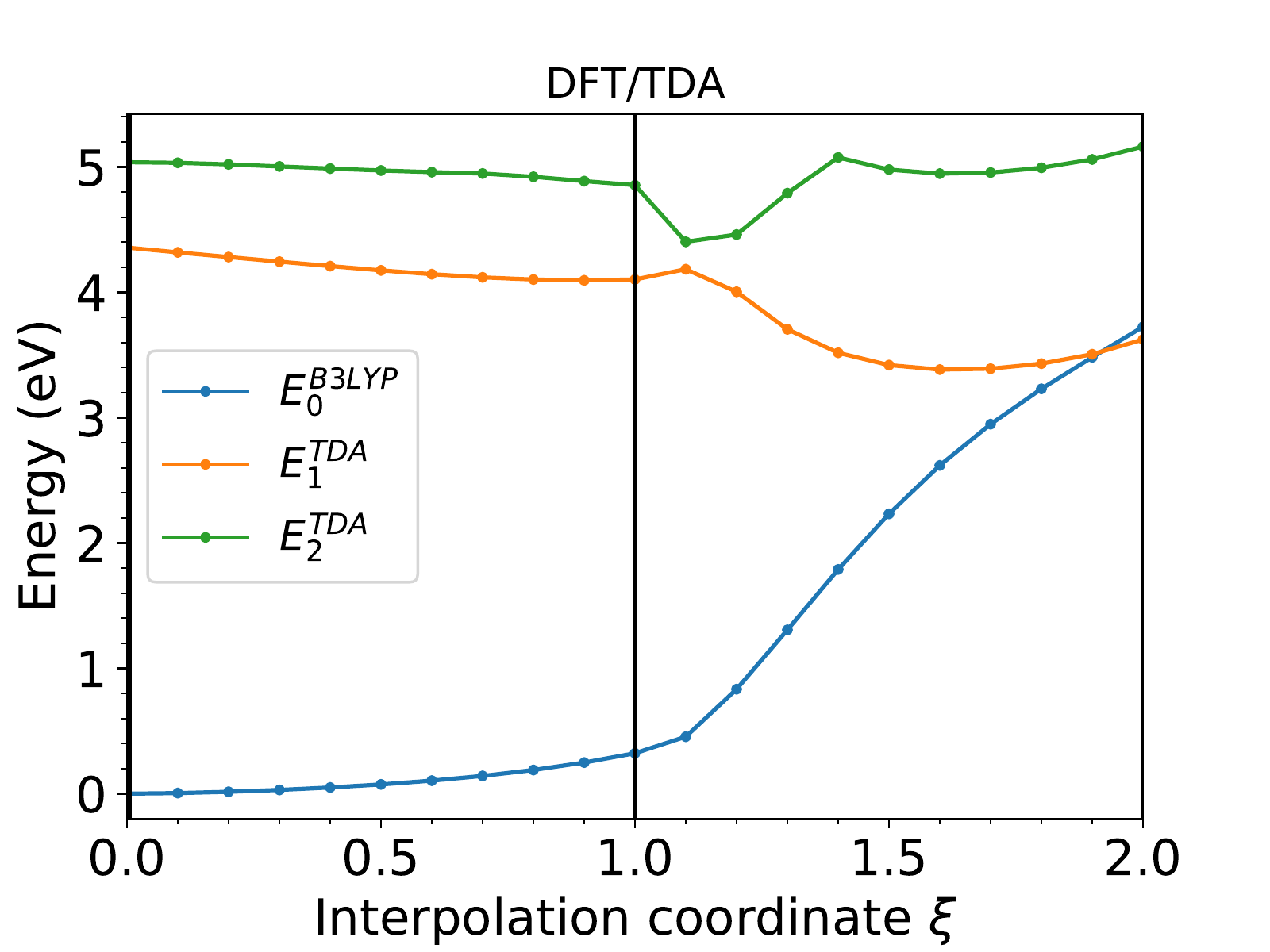}
    }
    \subfloat{
    \includegraphics[scale=0.52]{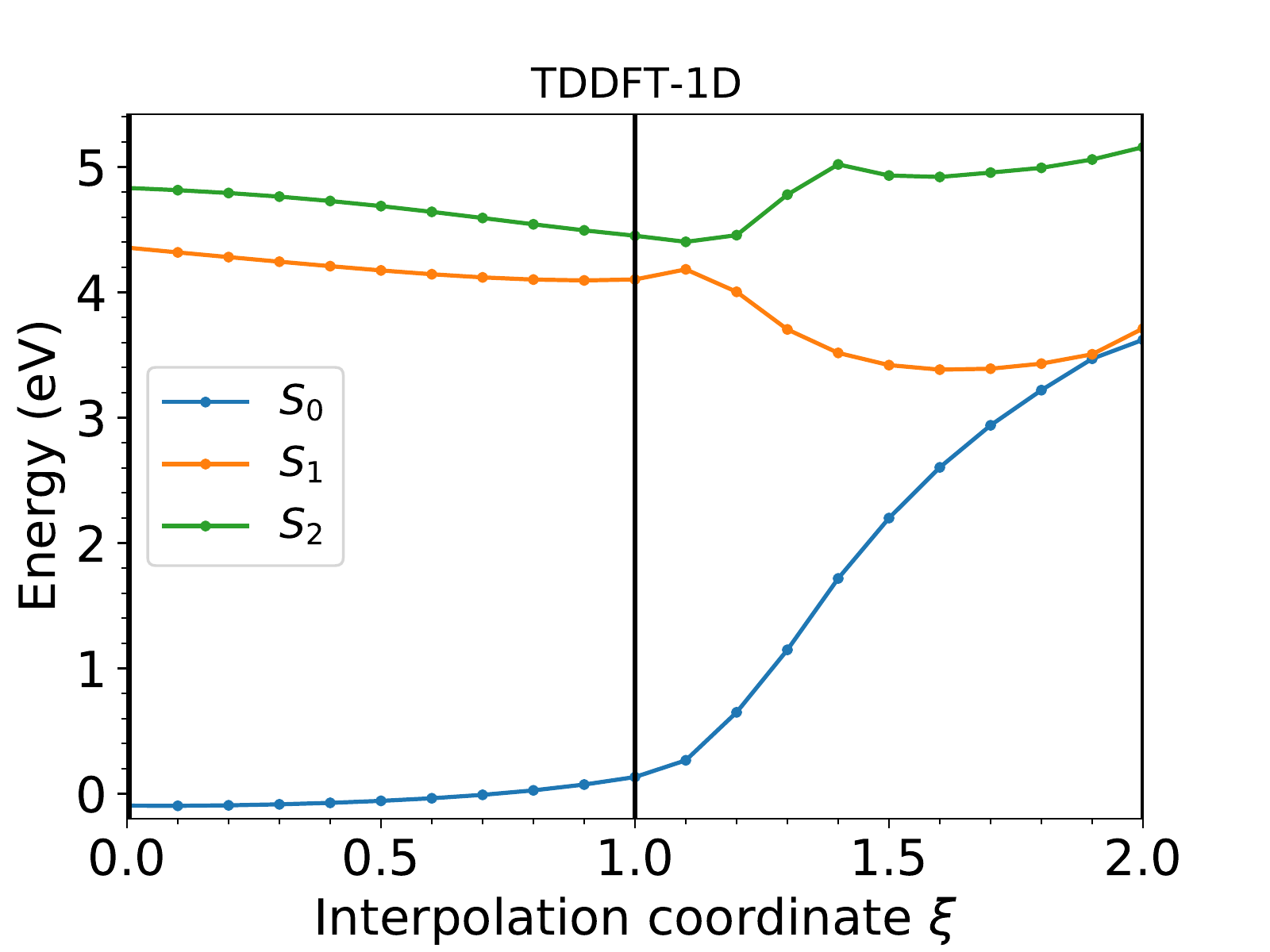}
    }
    \caption{Potential energy curves for the $S_0,S_1$ and $S_2$ states of the bithiophene molecule along the linear interpolation coordinate $\xi$ calulated using B3LYP/TDA (left) and TDDFT-1D (right). Here $\xi=0$ corresponds to the the $S_0$ minimum geometry, $\xi=1$ corresponds to the $S_1$-min-$a$ geometry, and $\xi=2$ to the $S_1$-min-$b$ geometry. For TDDFT, there is a conical intersection around $\xi = 1.8$; according to TDDFT-1D, this conical intersection becomes an avoided crossing. See Fig. \ref{fig:thiophene_zoom} for more information.}
    \label{fig:thiophene_energy_full}
    \end{figure}

In Fig.~\ref{fig:thiophene_energy_full}, potential energy curves are plotted along a linear interpolation coordinate starting from the $S_0$ minimum structure to the $S_1$-min-$a$, and continuing on to the open ring $S_1$-min-$b$. In Fig.~\ref{fig:thiophene_energy_full}a, we plot data for DFT/TDA and in \ref{fig:thiophene_energy_full}b we plot data for TDDFT-1D.  The vertical excitation energies to $S_1$ at the Franck Condon point for the B3LYP/TDA, TDDFT-1D and CASPT2 methods are 4.36 eV,4.45 eV, and 4.51 eV\cite{thiophene_main} respectively. These energies agree reasonably well with the experimental value of 4.24 eV\cite{thiophene_expt} (from an absorption spectrum at room temperature for the molecule in the gas phase).

\begin{figure}  
\includegraphics[scale=0.9]{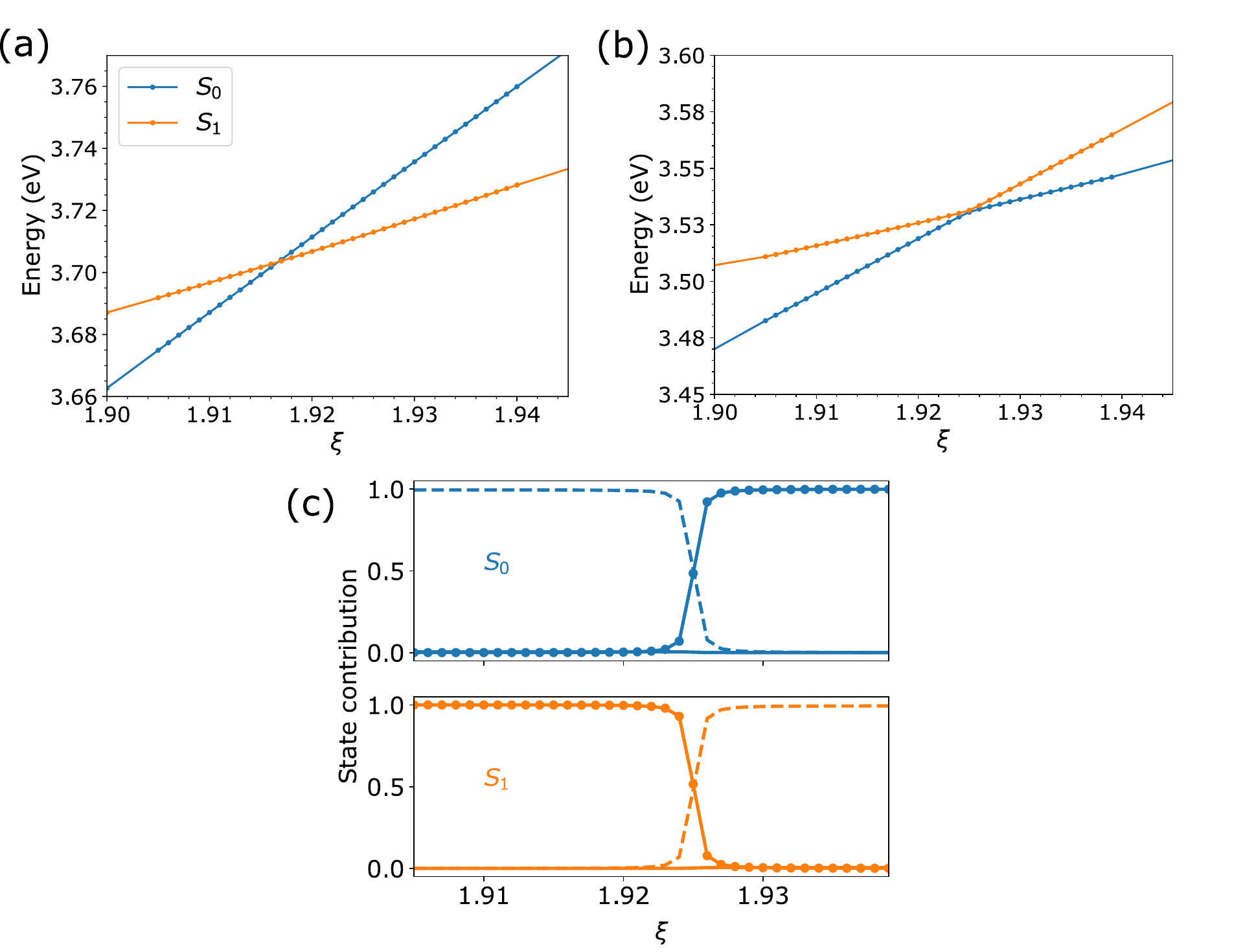}
    \caption{A zoom in on the $S_1/S_0$ potential energy curves plotted in Fig \ref{fig:thiophene_energy_full}. (a) The B3LYP/TDA calculation predicts a spurious crossing of the $S_1/S_0$ states, whereby the B3LYP state can sometimes have an energy above the TDDFT state. This behavior is corrected in the TDDFT-1D calculation (b), where the $S_0$ state always resides below the $S_1$ state.  In (c) the contribution of the KS determinant (dashed line), singly excited configurations (solid circles/line) and double (solid line) to each of the TDDFT-1D $S_0, S_1$ wave functions is plotted. The switching of the KS determinant and the singly excited configurations in the $S_0$ wave function at the crossing point is facilitated by the configuration interaction approach of TDDFT-1D.}
    \label{fig:thiophene_zoom}
\end{figure}

According to Fig.~\ref{fig:thiophene_energy_full}, both B3LYP/TDA and TDDFT-1D predict a $S_1/S_0$ crossing as we interpolate between $S_1$-min-$a$ and $S_1$-min-$b$. 
However, this crossing is unphysical for the B3LYP/TDA: because we employ a restricted DFT calculation, the $S_1$  (TDA) energy drops below the $S_0$ (DFT) energy around $\xi=1.9$, corresponding to a negative excitation energy.   In Fig.~\ref{fig:thiophene_zoom}, we zoom in on this spurious behavior. Note that for such TDDFT/DFT crossings, it is well known that the topology of an $S_1/S_0$ crossing is incorrect; the branching plane has the wrong dimensionality\cite{tddft_failure} (whether or not we use a restricted or unrestricted KS scheme).  Finally, note that all of these incorrect features are fixed up (at least qualitatively) using the TDDFT-1D calculation; for the present case, including one double introduces a small $S_1$-$S_0$ gap and shifts the location of the conical intersection slightly.  
\begin{figure}[ht]
    \centering
    \includegraphics[scale=0.35]{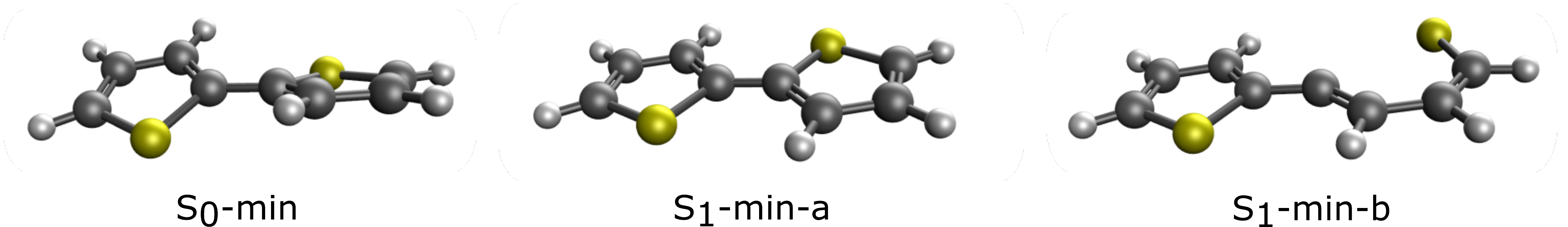}
    \caption{The structures of the optimized bithiophene molecule from Ref.\citenum{thiophene_main}. The $S_0$ minimum geometry is calculated using B3LYP/6-311G** and the two $S_1$ minima, $S_1$-min-$a$ and $S_1$-min-$b$, are calculated using CASSCF/6-31G*. }
    \label{fig:thiophene_geom}
\end{figure}
    

\subsection{LiF: A Test Case Balancing Accurate Energies and Smooth Crossings }
\begin{figure}
    \centering
    \includegraphics[scale=0.77]{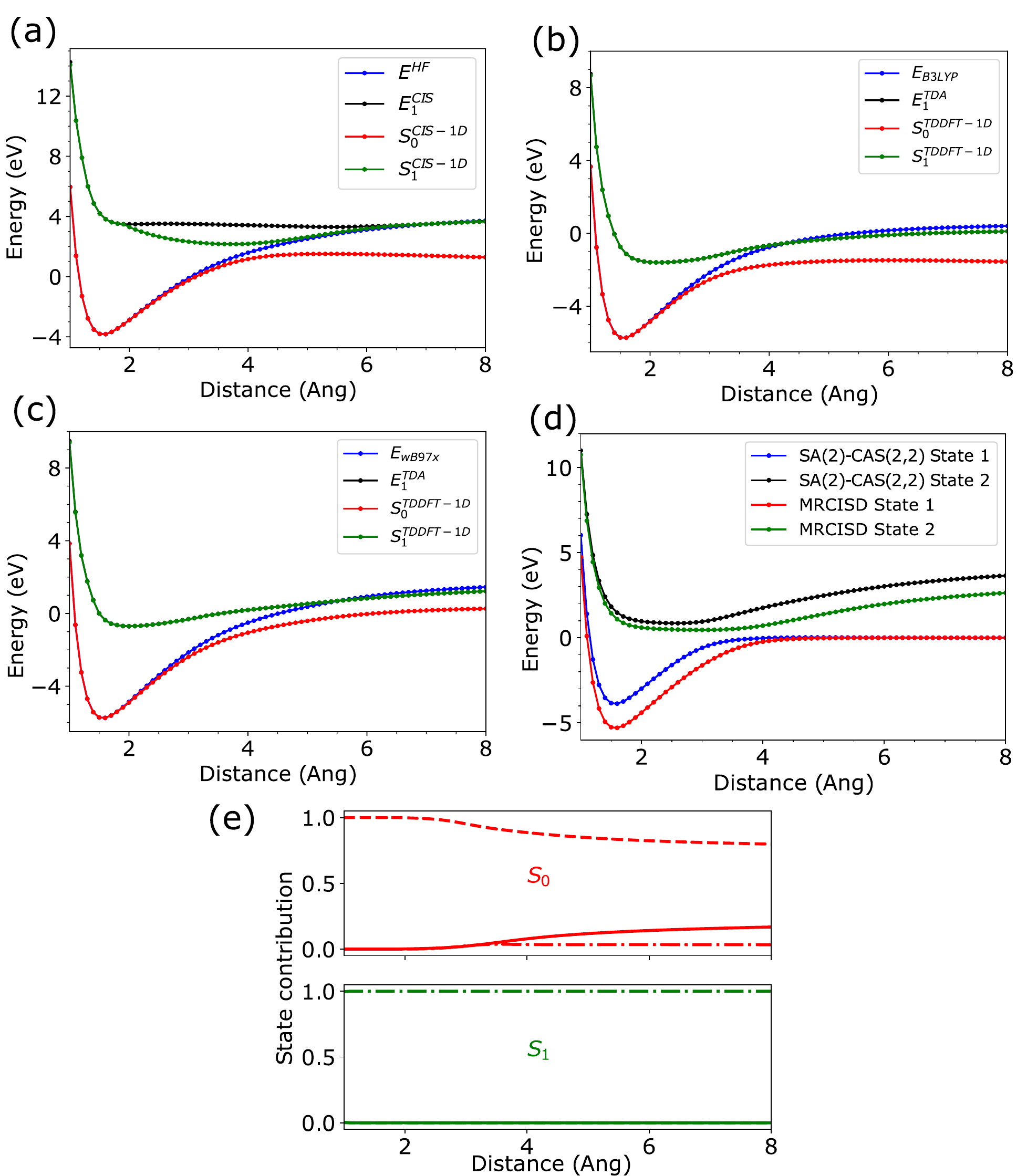}
    \caption{Potential energy curves of the lowest two singlet states for the dissociation of LiF molecule calculated using (a) RHF/CIS and CIS-1D, (b) B3LYP/TDA and TDDFT-1D, (c) $\omega$B97x/TDA and TDDFT-1D, and (d) SA(2)-CASSCF(2,2) and MRCISD. All the calculations were performed using the 6-31G* basis set. In (e) the contributions of the KS determinant (dashed line), singly excited configurations (dash-dot line) and double (solid line) to each of the TDDFT-1D/B3LYP $S_0,S_1$ wave functions are plotted. For the $S_0$ state, the  contribution of the double rises as the molecule dissociates and moves away from its equilibrium geometry.}
    \label{fig:lif_all}
    
\end{figure}
\begin{figure}
    \centering
    \includegraphics[scale=0.7]{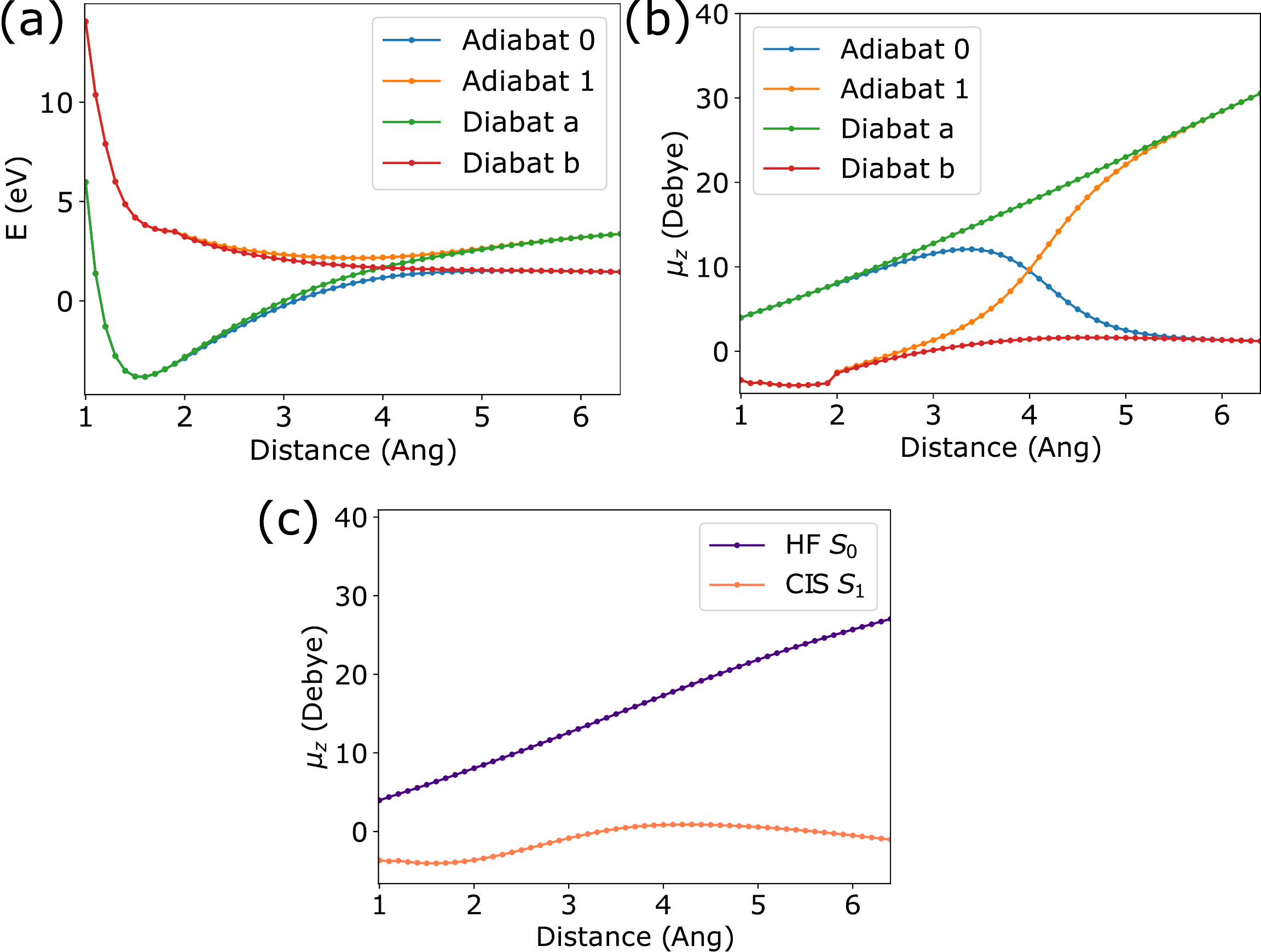}
    \caption{(a) CIS-1D Energy $E$ and (b) dipole moments for the lowest two adiabatic states (the $S_0$ and $S_1$ states) and diabatic states for LiF. In (c) the dipole moment of the lowest two adiabatic states from RHF/CIS calculation is plotted as a function of LiF bond distance. Note that while the CIS-1D adiabats show a crossing in their dipole moment (indicating a change in their charge character), the RHF/CIS adiabats do not cross, and predict the wrong charge character as LiF dissociates.}
    \label{fig:lif_diabat}
\end{figure}

Our last example is the dissociation curve of the LiF molecule, for which restricted Hartree-Fock (RHF) solution is known to be inadequate as far as dissociating into neutral Li and F fragments is concerned.
The ground state RHF curve leaves the Li and F atoms in a closed shell state with some ionic character;  this state becomes degenerate with the first excited CIS state as LiF dissociates. Although LiF is reasonably small, LiF represents a difficult case for CIS-1D/TDDFT-1D calculations because, at long distances, the true ground state is very different from the RHF solution; dissociating LiF is not like twisting the double bond of PYCM, whereby the doubles correction is important only at various intermediate angles (close to \ang{90}). Moreover, and most importantly, it is well known that most DFT functionals vastly underestimate the energy of charge-transfer states\cite{ct_failure_1_1,ct_failure_1_3,ct_failure_1_4,ct_failure_1_5}, and so one might wonder how a functional like B3LYP will perform when a doubly excited configuration is included in a semi-empirical Hamiltonian.  These questions will be addressed in Figs.~\ref{fig:lif_all}-\ref{fig:lif_diabat},
where we plot our results versus accurate
MRCISD results (that are based on a SA(2)-CASSCF(2,2) reference) obtained using the OpenMolcas\cite{openmolcas} software. See Fig.~\ref{fig:lif_all}d.
Furthermore, to make our analysis more explicit, in Table~\ref{tab:lif_quantities} we list data for the bond-dissociation energy, vertical excitation energy at the equilibrium geometry,  the effective diabatic coupling (at the crossing point), and the crossing point location. The dissociation energy is calculated as the energy difference between the minimum energy geometry and the energy at \SI{8.0}{\angstrom}.

We begin with the HF case.  In Fig.~\ref{fig:lif_all}a, we plot RHF vs CIS-1D potential energy curves.  We find that, by including one double only, the CIS-1D method is able to substantially correct for the static correlation errors and dissociate the molecule.  We plot the ground state $S_0$ and first excited state $S_1$ energy curves. Admittedly, the CIS-1D ground state curve is not completely size-consistent\cite{Teh2019}; the $S_0$ curve does not reach the sum of the individual Li and F energies (here, energy 0) when the molecule dissociates. Nevertheless, qualitatively, the behavior of the ${S_0}$ and ${S_1}$ states is correct and far improved over RHF. CIS-1D captures a reasonably accurate  dissociation energy.  Note that the CIS-1D vertical excitation energy is identical to the CIS vertical excitation energy (as we would normally prefer) -- but both are larger than the MRCISD result. See Fig.~\ref{fig:lif_all}.

Although not plotted here, note that while the spin contaminated unrestricted Hartree-Fock (UHF) solution does recover the correct zero energy dissociation limit (i.e. the method is size-consistent), as is well known, the UHF $S_0$ curve shows a kink\cite{dutoi2006self} at the point where RHF and UHF solutions diverge. Furthermore, UCIS energies computed on top of a UHF reference are wildly discontinuous. By contrast, the CIS-1D solution is spin-pure, smoother, and at least qualitatively correct.

Next, in Fig.~\ref{fig:lif_all}b, a B3LYP/TDDFT-1D calculation is compared with that of B3LYP/TDA. First, note that the vertical excitation energy of B3LYP is too small; see Table~\ref{tab:lif_quantities}. Second, note that the restricted KS method is unable to dissociate the molecule correctly and actually shows an unphysical crossing with the TDA state around \SI{4.3}{\angstrom}. By contrast, according to TDDFT-1D, the ground state $S_0$ and excited state $S_1$ solutions are well separated and dissociate with the correct asymptotic behavior.
Third, however, note that although the B3LYP-1D potential energy curve is qualitatively correct, the TDDFT-1D method does not solve the problems of B3LYP; the excitation energy is still too small and, likely because of the charge transfer problem, the dissociation energy is now too small. Unfortunately, it is clear that the 1D method cannot solve the charge-transfer problem in TDDFT\cite{ct_failure_1_1,ct_failure_1_3,ct_failure_1_4,ct_failure_1_5}.


With the facts above in mind, we turn to a DFT functional ($\omega$B97x/TDA ) with built-in long-range exchange and reexamine the dissociation curve for LiF. Like B3LYP, the $\omega$B97x/TDA curves show an incorrect crossing of the ground $S_0$ and excited $S_1$ state; however, the excitation energy for $\omega$B97x/TDA  is not as low as it is for B3LYP. If we now apply the 1D ansatz with a scaling factor of $\alpha=0.5$, $\beta=0.75$, we find that (as desired) the vertical excitation energy does not change, but the ground state $S_0$ curve no longer crosses over the $S_1$. Thus, again, the 1D correction is an improvement over the standard TDA approach.  That being said, in truth, the $S_0$ curve does not have the correct behavior--it does not flatten out entirely as it should by \SI{8}{\angstrom} . In other words, including a 1D configuration is far from a sinecure for all of the problems of TDDFT.  As a side note, we mention that empirically, we have found that different DFT functionals will benefit most with different scaling parameters. In other words, our results for $\omega$B97x/TDDFT-1D  are not optimized by choosing $\alpha=0.5$, $\beta=0.75$; a better semiempirical scheme can likely be found by benchmarking to ascertain the optimal parameters for each DFT functional.

\begin{table}[htbp]
\caption{Quantities from the potential energy curves for LiF: bond dissociation energy ($D_e$), vertical excitation energy at equilibrium geometry ($\Delta E_{vert}$), minimum energy separation between the curves which occurs at the diabatic curve crossing ($V_c$) and the interatomic distance at which crossing happens ($x_c$) } 
\label{tab:lif_quantities}
\begin{tabular}{lrccc}
\hline
Method         & $D_e$ (eV) & $\Delta E_{vert}$ (\si{\eV}) & $V_c$ (\si{\eV}) & $x_c$ (\si{\angstrom}) \\ \hline
HF/CIS         & $>7.54$  & 7.64                    & NA  & NA   \\
CIS-1D         & $5.11$  & 7.64                    & 0.92  & 4.3   \\
B3LYP/TDA      & $>6.13$  & 4.59                    & NA  & NA   \\
TDDFT-1D/B3LYP       & 4.18  & 4.59                    & 1.05  & 3.6   \\
$\omega$B97x/TDA      & $>7.19$  & 5.38                    & NA  & NA   \\
TDDFT-1D/$\omega$B97x  & $>6.01$  & 5.38                    & 0.86  & 6.1   \\
SA(2)-CAS(2,2) & 3.87  & 5.35                    & 1.43  & 3.3   \\ 
MRCISD         & 5.30  & 6.39                    & 0.95  & 4.0 \\ \hline
\end{tabular}
\end{table}


Lastly, let consider diabatization and charge transfer dynamics.  One of the biggest advantages of a multireference method is the capacity to generate diabatic states for $S_1/S_0$ crossings. Since CIS-1D has some multi-reference character -- e.g.,  the states produced from CIS-1D (TDDFT-1D) are linear combinations of the reference HF (KS) wave function, the singly excited configurations, and one doubly excited configuration -- one would hope that CIS-1D can produce meaningful diabats.  After all, the  method produces ground and excited states on a reasonably equal footing. This state of affairs stands in contrast to any diabatization scheme based on mixing the HF ground state and CIS states. In the latter case, one expects that all results will be far less meaningful on account of Brillouin's theorem such that, e.g., there is no guarantee of smooth diabatic surfaces.

In Fig.~\ref{fig:lif_diabat}, we  plot CIS-1D diabats as produced using the Boys localization method\cite{boys_subotnik2008}. Applying Boys diabatization to CIS-1D (TDDFT-1D) states is fairly straightforward
and requires only the dipole moment matrix elements, which are calculated explicitly in the Appendix.  In Fig.~\ref{fig:lif_diabat}a, we plot the two lowest CIS-1D adiabatic and diabatic  energies for the LiF molecule; in Fig.~\ref{fig:lif_diabat}b, we plot the relevant dipole moments in the avoided crossing region. Clearly, CIS-1D is able to reproduce smooth, qualitatively correct diabatic states. The dipole moment of the CIS-1D ground state is ionic at equilibrium bond distances but
correctly vanishes as the LiF molecule dissociates into neutral fragments.  The CIS-1D diabatic states have fixed charge character and correctly interpolate between the $S_0$ and $S_1$.  Note that the RHF and CIS dipole moments, plotted in Fig.~\ref{fig:lif_diabat}c, are (of course) completely invalid. From this data, we may conclude that CIS-1D (and TDDFT-1D provided we use a range-corrected functional and we can ascertain optimal parameters) can describe charge transfer dynamics -- at least qualitatively and sometimes in fact quantitatively.

\section{Discussion}\label{discussion}
\subsection{Convergence}
\begin{figure}[htbp]
    \centering
    \includegraphics[scale=0.65]{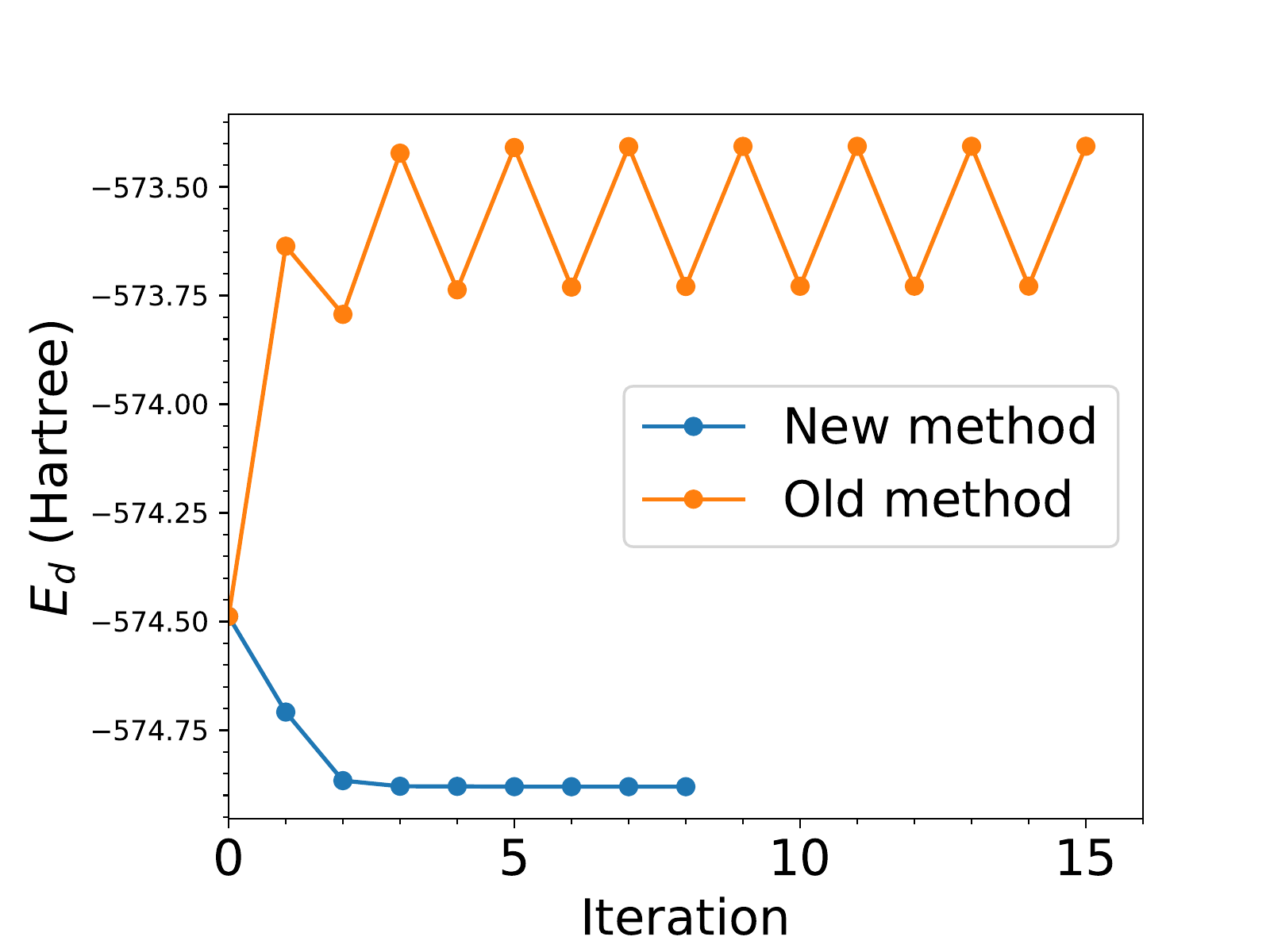}
    \caption{The energy of the doubly excited state $E_d$ at each iteration step of the minimization process for PYCM molecule at $\tau=\ang{40}.$ The old method of minimization in Ref.\citenum{Teh2019} is unable to find the optimized frontier orbitals as $E_d$ oscillates around a value that is higher than that obtained from the initial set of orbitals. In contrast, the new method outlined in this manuscript finds a solution that converges fairly quickly to the correct minimum. }
    \label{fig:newtonVsSCf}
\end{figure}

The data above has demonstrated that, when converged and with the proper semiempirical parameters,  TDDFT-1D can predict a large qualitative change in practical electronic structure calculations by including one double configuration.  Having found such a result, let us now show how convergence can be difficult and why the present algorithm is necessary.  In Fig.~\ref{fig:newtonVsSCf}, we demonstrate that the previous minimization algorithm (described in Ref.~\citenum{Teh2019}) fails to converge for the PYCM molecule.  In particular, note that Newton-Raphson minimization converges in 8 iterations, while the old method at first {\em increases in energy} and eventually simply oscillates back and forth along the incorrect asymptote.  We have also found that the algorithm in Ref.~\citenum{Teh2019} can fail to converge to the correct set of orbitals for certain LiF geometries.  In general, by coordinating the occ-occ and virt-virt rotations through the Hessian $\mathcal{H}$, the Newton-Raphson method is clearly quite a few steps ahead with regards to the stability of optimization.

At this point, the key item remaining is the question of efficiency and computational time: how expensive is it to construct the orbital Hessian above in Sec.~\ref{theory}?
To answer this question, at the moment it is essential to distinguish between CIS-1D and TDDFT-1D.  First, we address CIS-1D. 
For the present implementation, the cost of the algorithm comes out to be 1.9 s wall time per iteration (for the case of the LiF molecule), whereas the algorithm in Ref.~\citenum{Teh2019} takes 0.5 s per iteration for the same molecule -- but note that the latter algorithm never converges.
Overall, the algorithm from Ref.~\citenum{Teh2019} requires  the diagonalization of two matrices (one of size $N_o\times N_o$, another of size $N_v \times N_v$).  By contrast, the present algorithm requires the inversion of one matrix (of size $N \times N$).  Here, $N_o$ is the number of occupied orbitals, $N_v$ is the number of virtual orbitals, and $N = N_o + N_v$ is the total basis size.
Thus, the present algorithm should be some constant factor more expensive than the previous algorithm (per iterative step). Note that the present algorithm also requires computing three sets of density matrix times electron repulsion integrals (so-called J/K subroutines), whereas the previous algorithm required only one such call. In the end, one might expect that, per iteration, the present algorithm will be three times as expensive as the previous approach -- though requiring far fewer iterations and also converging in a far more robust fashion.
\par The total time for running CIS-1D is summarized in Table~\ref{tab:timings}. For LiF (two atoms),  a CIS-1D run on a single thread requires \SI{0.57}{\second}, of which \SI{0.17}{\second} is needed for the step to minimize the energy of the double excitation configuration. For the medium-sized molecule PYCM (28 atoms), the total wall time is 81.81 s running on 32 threads, of which \SI{18.42}{\second} are spent on $E_d$ minimization. The remaining \SI{63.39}{\second} are spent on diagonalization. This large amount of time implies that our current implementation is far  from optimal since the corresponding CIS calculation requires only \SI{9.2}{\second}. Since the configuration interaction Hamiltonian in CIS-1D and CIS differ by only 2 in dimensionality, further improvements are clearly possible and will be addressed in the future. 

\begin{table}[ht]
\caption{Timing for CIS-1D calculations for PYCM and LiF molecules. The CIS-1D total time is composed of two steps: $E_d$ minimization step and diagonalization of the configuration interaction hamiltonian. The timing for a CIS calculation is given as a reference. For a small molecule, the cost of CIS-1D is comparable to CIS while the cost for CIS-1D is ten-fold for a larger molecule with the current implementation.  Future improvements are clearly possible, as currently the bulk of the time is spent in the diagonalization step (even though the number of basis functions in  CIS-1D  is only two more than that in CIS).  }
\label{tab:timings}
\centering
\begin{tabular}{lcc}
\hline
\textbf{PYCM} (28 atoms)    &              &               \\ \hline
                   & CPU time (s) & Wall time (s) \\ \hline
CIS                & 183.73       & 9.2           \\
CIS-1D Diagonalization     & 1218.10 & 63.39         \\
$E_d$ minimization & 522.83       & 18.42         \\
CIS-1D Total       &  1740.93      & 81.81               \\\hline
\textbf{LiF} (2 atoms)      &              &               \\ \hline
                   & CPU time (s) & Wall time (s) \\ \hline
CIS                & 0.17         & 0.17          \\
CIS-1D  Diagonalization            &   0.37 & 0.37      \\
$E_d$ minimization & 0.17         & 0.17          \\ 
CIS-1D Total       & 0.54         & 0.54        \\\hline
\end{tabular}
\end{table}

Next, we turn to TDDFT-1D, for which the bottleneck for the minimization algorithm is the calculation of the matrix elements of the exchange-correlation kernel in the adiabatic approximation $f^{xc}(\vb{r},\vb{r'}) = {\pdv[2]{E_xc}{\rho(\vb{r})}{\rho(\vb{r'})}}$ (Eq. \ref{eq:dft_ed_diff}).  In particular, in order to minimize a doubly excited configuration in a molecular orbital basis, we require matrix elements of the form $f^{xc}_{ih,jh}, f^{xc}_{ih,al},$ and $f^{xc}_{al,bl}$ .  Unfortunately, within 
current computational codes (e.g. Q-Chem\cite{qchem}), such matrix elements are not readily available and we have currently implemented a painful (and slow) approach towards calculating these $\sim N^2$ matrix elements.  Future work will necessarily need to construct these matrix elements in a timely and efficient fashion in order for the TDDFT-1D approach to be fast and competitive. This project is now ongoing.

\subsection{$S_1/S_0$ crossings a function of Scaling Parameters}

\begin{figure}
     \includegraphics[scale=0.8]{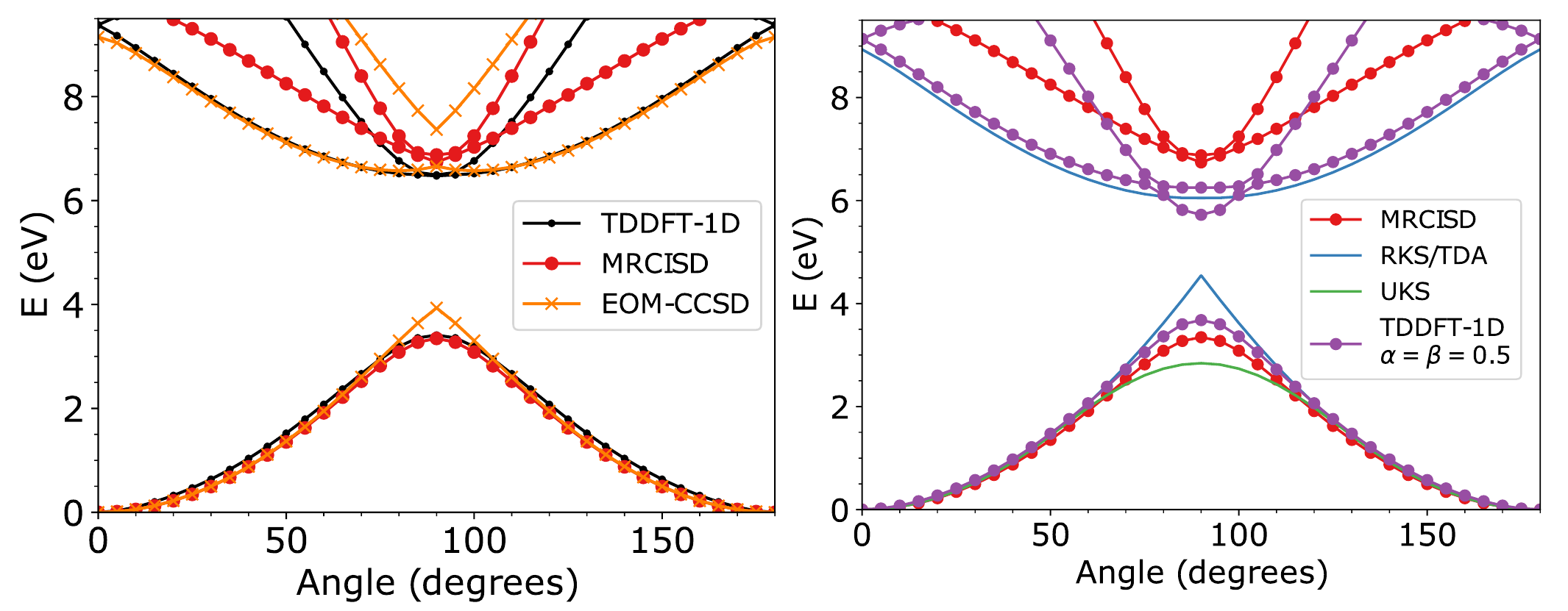}
     \caption{Potential energy curves for ethylene molecule computed using TDDFT-1D, MRCISD, and EOM-CCSD (left). On the right, we show an example TDDFT-1D calculation with a different set of scaling parameters ($\alpha=0.5$, $\beta=0.5$). Reference curves calculated from RKS/TDA and UKS are also shown.  Around the crossing point when the molecule is twisted to \ang{90}, CCSD and RKS show unphysical cusps. The TDDFT-1D ground state curve calculated with BHHLYP functional is in good agreement with that calculated from MRCISD. Using non-optimal scaling factors, however, can make undesirable shifts in the energy levels of various states.}
    \label{c2h4_fig}
\end{figure}

In Sec.~\ref{benchmark_section} above, we have benchmarked the  TDDFT-1D algorithm using a choice of parameters ($\alpha=0.5$ and $\beta = 0.75$) that was designed to be an empirical compromise between (i) undisturbed vertical excitation energies and (ii) smooth $S_1/S_0$ crossings.  Obviously, goal (i) requires that the $\alpha$ and $\beta$ parameters be small.  With this in mind, for the sake of completeness, we will now address how the choice of parameters affects goal (ii). To do so,      
we revisit the potential energy curves for ethylene molecule for the coordinate of rotation along the double bond, as presented in Ref.~\citenum{Teh2019}. First, in Fig.~\ref{c2h4_fig}, we compare the curves obtained using MRCISD starting from a CASSCF(2,2) wave function with the TDDFT-1D curve calculated using BHHLYP functional using the 6-31G* basis set. We also show the curve obtained from EOM-CCSD which is known to show a sharp cusp at the midpoint of the bond rotation. It is clear that TDDFT-1D with optimal scaling parameters behaves reasonably well.

Second, in Fig.~\ref{c2h4_fig}(b), we compare the curves for another version of TDDFT-1D ($\alpha=0.5$, $\beta = 0.5$).  We find that a non-optimal choice of scaling parameters leads to problems in the potential energy curves. In particular, for the $\alpha=0.5$, $\beta = 0.5$ data, the gap is far too small and the $S_2$ energy is not equal to the $S_1$ energy at \ang{90} as it should be.  For completeness, we also include the RKS and UKS ground states. As is well known, the RKS ground state around \ang{90} is no longer smooth, and the $S_2$ TDA excited state with the correct character cannot be obtained at this geometry. With an unrestricted ansatz, the ground state UKS curve is smooth but the method significantly underestimates the barrier height. Moreover, TDA excited states from a UKS reference are strongly spin contaminated and hence not shown in the plot.

In the end, $\alpha=0.5$, $\beta = 0.75$ would appear to be a very good compromise set of parameters that fulfill the goals that we set out. These parameters predict excitation energies that are as accurate as TDDFT/TDA over a reasonably large set of molecules while producing smooth potential energy curves in situations that are not easily handled by DFT/TDA.

\begin{figure}[htbp]
    \centering
    \label{fig:lif_wb97x_compare}
\end{figure}

\section{Conclusion}
We have shown that Newton-Raphson optimization is stable as far as finding the optimal doubly excited configuration for the CIS-1D and the TDDFT-1D method. We have also shown that, with an optimal choice of scaling parameters for the TDDFT-1D functionals, one can find a meaningful balance between accurate vertical excitation energies and smooth $S_1/S_0$ crossings, such that these configuration interaction Hamiltonians can successfully recover electronic potential energy surfaces states corresponding to bond making/breaking processes.  Furthermore, one can use such approaches to produce meaningful diabatic states through adiabatic-to-diabatic transformations (e.g. the Boys diabatization method). 

In the present paper, we have worked almost exclusively with the B3LYP functional.  Clearly, more benchmarking work will be necessary if one wishes to  generate a semi-empirical TDDFT-1D ansatz for different DFT functionals. Nevertheless, if we are prepared to implement a semi-empirical configuration interaction Hamiltonian on top of a DFT ansatz, the goal of a reasonably accurate algorithm (that does not fail as far as $S_1/S_0$ crossings and is suitable for dynamics) does seem within reach using the present practical formalism.
Moreover, once the necessary matrix elements of $f^{xc}$ are computed efficiently, the present method should be applicable to quite large molecules and act as an alternative to spin-flip methods\cite{sasfcis-herbert,sfdft2}. Ultimately, if one can successfully apply the techniques of Ref.~\citenum{teh2020analytic} for CIS-1D analytical derivatives\cite{furche_deriv_coupling} to the TDDFT-1D method, simulating nonadiabatic chemical dynamics\cite{faraji_dynamics_study,matsika_dynamics} within such an ansatz should be on the near horizon.

\begin{acknowledgements}
This work was supported by the U.S. Air Force Office of Scientific Research (USAFOSR) AFOSR Grants No. FA9550-18-1-0497 and FA9550-18-1-0420. 
\end{acknowledgements}

\section*{Data Availability Statement}
The data that support the findings of this study are available from the corresponding author upon reasonable request.

\appendix
\section{Dipole moments of CIS-1D/TDDFT-1D states}
Here, we provide the relevant equations for obtaining dipole moments in a CIS-1D calculation.  A CIS-1D state
is a linear combination of its basis functions: 
\begin{equation}
\ket{\psi_{CIS-1D}} = c_0\ket{\phi_{0}} + \sum_{ia}\frac{1}{\sqrt{2}}c_i^a(\ket{\phi_i^a}+\ket{\phi_{\bar{i}}^{\bar{a}}}) + c_d\ket{\phi_{h\bar{h}}^{l\bar{l}}}.    
\end{equation}
 Note that the ground state HF wave function is restricted, and the singly excited configurations are singlets. The dipole moment between two  CIS-1D states is then given by
\begin{align}
    \mel{\psi^I}{X}{\psi^J} &= c_o^Ic_0^JX_{0}+\sum_{ia}{\sqrt{2}}(c_o^Ic_i^{aJ}+c_o^J{c_i^{aI}})X_{ia} \nonumber
    \\&\quad+\sum_{ia}c_i^{aI}c_i^{aJ}X_{0}+\sum_{iab}c_i^{aI}c_i^{bJ}X_{ab}-\sum_{ija}c_i^{aI}c_j^{aJ}X_{ij} \nonumber \\ 
    &\quad+\sum_{ia}\sqrt{2}(c_h^{lI}c_d^J+c_h^{lJ}c_d^I)X_{hl} + c_d^Ic_d^J(X_{0}+2X_{ll}-2X_{hh}).
    \label{dipole_elems}
\end{align}
Here $X_0$ is the ground state dipole moment, and $X_{pq}$ are the elements of the dipole moment matrix.

 


\nocite{*}
\bibliography{optOrb}

\end{document}